\documentclass[english,twocolumn]{emulateapj}
\usepackage{amsmath}
\usepackage{mathtools}

\usepackage{color}

\shorttitle{Circumbinary Planets}
\shortauthors{Li, Holman \& Tao}

\begin{document}


\title{Uncovering Circumbinary Planetary Architectural Properties from Selection Biases}
\author{Gongjie Li \altaffilmark{1}, Matthew J. Holman \altaffilmark{1} and Molei Tao \altaffilmark{2}}
\affil{$^1$ Harvard-Smithsonian Center for Astrophysics, The Institute for Theory and
Computation, \\60 Garden Street, Cambridge, MA 02138, USA}
\affil{$^2$ School of Mathematics, Georgia Institute of Technology, 
\\686 Cherry Street, Atlanta, GA 30332, USA}
\email{mtao@gatech.edu}



\begin{abstract}
The new discoveries of circumbinary planetary systems shed light on the understanding of planetary system formation. Learning the architectural properties of these systems is essential for constraining the different formation mechanisms. We first revisit the stability limit of circumbinary planets. Next, we focus on eclipsing stellar binaries and obtain an analytical expression for the transit probability in a realistic setting, where finite observation period and planetary orbital precession are included. Then, understanding of the architectural properties of the currently observed transiting systems is refined, based on Bayesian analysis and a series of hypothesis tests. We find 1) it is not a selection bias that the innermost planets reside near the stability limit for eight of the nine observed systems, and this is consistent with a log uniform distribution of the planetary semi-major axis; 2) it is not a selection bias that the planetary and stellar orbits are nearly coplanar ($\lesssim 3^\circ$), and this together with previous studies may imply an occurrence rate of circumbinary planets similar to that of single star systems; 3) the dominance of observed circumbinary systems with only one transiting planet may be caused by selection effects; 4) formation mechanisms involving Lidov-Kozai oscillations, which may produce misalignment and large separation between planet and stellar binaries, are consistent with the lack of transiting circumbinary planets around short-period stellar binaries, in agreement with previous studies. As a consequence of 4), eclipse timing variations may better suit  the detection of planets in such configurations. 

\bigskip
\end{abstract}


\section{Introduction}

The exciting discoveries of circumbinary planetary systems can provide better understanding of planetary formation. To date, 11 transiting circumbinary planets have been discovered residing in nine planetary systems, including Kepler-16b \citep{Doyle11}, Kepler-34b and 35b \citep{Welsh12}, Kepler-38b \citep{Orosz12b}, Kepler-47b, 47c \citep{Orosz12a} and 47d (\citealt{Hinse15}, Orosz et al. in prep.), Kepler-64b \citep{Schwamb13, Kostov13}, Kepler-413b \citep{Kostov14}, Kepler-453b \citep{Welsh15} and Kepler-1647b \citep{Kostov15}. Many of them share interesting architectural features. For instance, the locations of the planets are mostly near the stability limit, the mutual inclinations between the planetary orbits and the stellar binary orbits are low, and the planets preferentially orbit around stars with long stellar orbital periods. 

The architectural properties of these systems reveal important clues to the origin of circumbinary planetary systems. For instance, the observed pile up of planets near the stability limit may indicate the dominance of disk migration, as the planets move to the instability limit via disk migration from their birth location, which is likely farther away \citep[e.g.,][]{Paardekooper12, Rafikov13, Marzari13, Pierens13, Kley14, Bromley15, Silsbee15}. In addition, the near coplanar configuration of the circumbinary planetary systems around closely separated binary stars is consistent with theoretical studies of the gravitational torque between the binary and the circumbinary disk, which produces the alignment \citep{Foucart13, Foucart14}. Moreover, the coplanarity is also consistent with the observed alignment of protoplanetary disk around the young binary stars \citep{Rosenfeld12, Czekala15, Czekala16}, and with the debris disk around short period binaries \citep{Kennedy12}. This may imply a primordial origin of the alignment of the planetary orbits. Note that 99 Herculis hosts a misaligned circumbinary debris disk, and the origin of the misalignment challenges the collisional and dynamical evolution of the system \citep{Kennedy12b}.

Some of the architectural features are caused by dynamical interactions, which play critical roles in the origin of the planetary systems. Hierarchical three-body system dynamics has been studied in the literature extensively. In the case when the inner binary contains of a test particle, the eccentricity and inclination of the inner binary can oscillate due to the perturbation of the outer object under the Lidov-Kozai mechanism \citep{Kozai62, Lidov62}. Including the octupole order of expansion (third power in the semi-major axis ratio), it has been shown that the inner orbit can change from a prograde orbit to a retrograde one and the eccentricity can be excited very close to unity \citep[e.g.,][]{Naoz11, Katz11, Li14}. In the case when the outer object is a test particle, which is more relevant to the circumbinary planets, the Lidov-Kozai oscillations disappear \citep{Migaszewski11, Martin16}. On the other hand, multiple equilibria exist when the mutual inclination is high \citep{Palacian06, Verrier09, Farago09, Doolin11}. In addition, the orbit of the test particle is not stable when it is very close to the binary. The stability limit have been obtained for a large parameter space \citep[e.g.,][]{Dvorak89, Holman99, Musielak05, Doolin11}, and outcomes of the unstable systems have been investigated \citep{Sutherland15, Smullen16}.

In addition to the dynamical effects, selection biases also influence the observed architectural properties. Thus, correcting selection biases is crucial when one extracts the architectural properties from the observed circumbinary systems. For instance, the detection limitation favors planets that are closer to the stellar binaries, and it is more likely to detect planets that are coplanar with the eclipsing stellar binaries using the transit method. Considering selection effects, \citet{Armstrong14} studied the abundances and properties of the circumbinary systems extensively using the approach of population synthesis. It was found that the occurrence rate of circumbinary planetary systems has a lower limit of $47\%$ if the mutual inclination between the planetary orbit and the stellar binary is isotropically distributed. This implies that the circumbinary systems are preferentially coplanar or can be formed much more easily than the single star systems. However, precession was neglected in the previous derivation of the occurrence rate, and it has been shown that precession plays an important role in the transit probability \citep{Schneider94, Martin15}. In particular, \citet{Martin15} shows that if one takes an infinite amount of time, the transit probability for a circumbinary planetary system is larger than that of a single star system, and the transit probability increases with mutual inclination. It is not realistic to consider an infinite amount of observation time, yet the transit probability over a finite observation time has not been derived analytically. Here, we revisit the transit probability to derive the transit probability in a finite observation time and include orbital precession for planets orbiting eclipsing binaries. Then, we correct selection biases using transit probabilities in order to obtain the architectural properties of the observed circumbinary planetary systems. This differs from \citet{Martin14}, who considered selection biases for planets orbiting both eclipsing and non-eclipsing binaries using a large number of synthetic systems based on N-body simulations. Specifically, they found that the pile-up of the planets near the stability limit is not due to selection biases, and the coplanarity of the systems may indicate either a high occurrence rate of circumbinary systems or that the coplanarity is not a selection effect. In this article, we focus on the observed systems, and we use Bayesian analysis to study the selection effects. For instance, the coplanarity is not degenerate with the occurrence rate of the circumbinary planetary systems this way.

This article is organized as the following: in \textsection \ref{s:sta}, we revisit the stability of the circumbinary planets including high mutual inclinations between the orbits of the planet and stellar binary. In \textsection \ref{s:tprob}, we provide an analytical expression for the transit probability in a finite observational period as a function of stellar and planetary orbital parameters. In the end (\textsection \ref{s:arc}), we study the circumbinary architecture corrected from selection biases using the transit probability. 
 
\section{Stability Limit as a Function of Mutual Inclination}
\label{s:sta}

The stability of the circumbinary systems has been studied \citep[e.g.,][]{Dvorak89, Holman99, Pilat03, Musielak05, Doolin11}. In particular, \citet{Doolin11} discussed many interesting features of the parameter space where the systems are stable, extending mutual inclination between the orbits of the stellar binary and the planet from $0^\circ$ to $180^\circ$, and including eccentric stellar binaries and different stellar mass ratios assuming the planet is massless. It has been found that the retrograde orbits are more stable than the prograde orbits, which is true in three-body problems in general. In addition, there exist striations of instability likely due to resonances between the stellar binary and the planet, and there are pinnacles and peninsulas of unstable regions for the non-librating and librating regions respectively, except when the stellar masses are equal. 

The critical semi-major axis within which the planet is unstable for the coplanar case was obtained by \citet{Holman99}, who performed a large number of numerical simulations that cover a wide range of stellar binary eccentricity and stellar mass ratio. For reference, $a_c$ is expressed as the following:
\begin{align}
		\label{eqn:sta}
a_c = & \Big[(1.60 \pm 0.04) + (5.10 \pm 0.05)e \\ \nonumber
		&+ (-2.22\pm0.11)e^2 + (4.12\pm0.09)\mu \\ \nonumber
		&+ (-4.27\pm0.17)e\mu + (-5.09\pm0.11)\mu^2 \\ \nonumber
		&+(4.61\pm0.36)e^2\mu^2\Big] a_b ,
\end{align}
where $e$ is the stellar binary eccentricity, $\mu = m_2/(m_1+m_2)$ is the stellar mass to total stellar mass ratio, and $a_b$ is the stellar binary semi-major axis. The expression is obtained by fitting the results of the numerical simulations, and the uncertainties are inherent from the fit. Note that this stability limit works only for the coplanar cases. Since discussions will be made on the misaligned cases in the next sections, we first illustrate the stability limit and the marginally stable parameter space as a function of the mutual inclination here.

\begin{figure}
\begin{center}
\includegraphics[width=3.3in, height=2.5in]{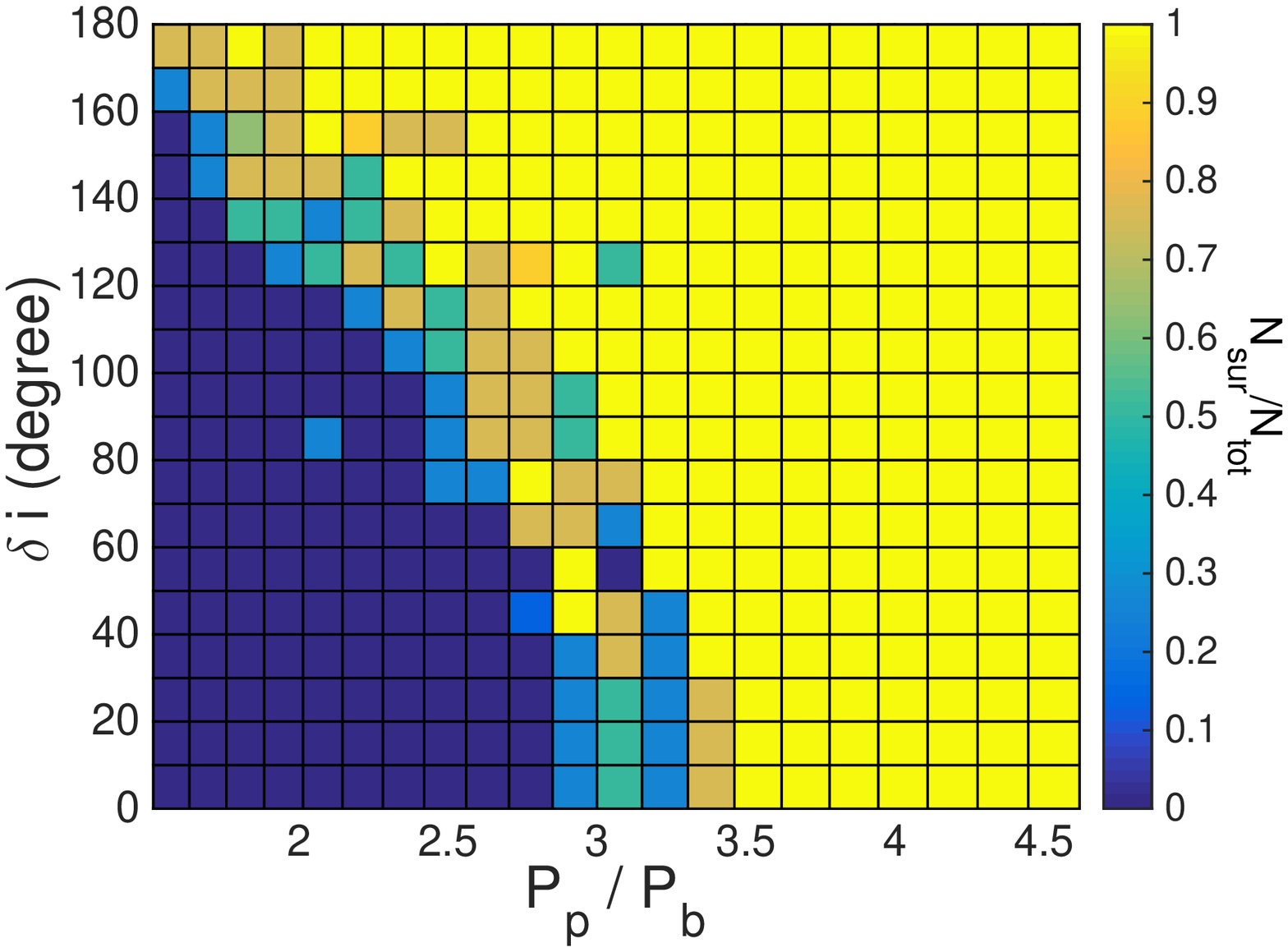} \\
\includegraphics[width=3.3in, height=2.5in]{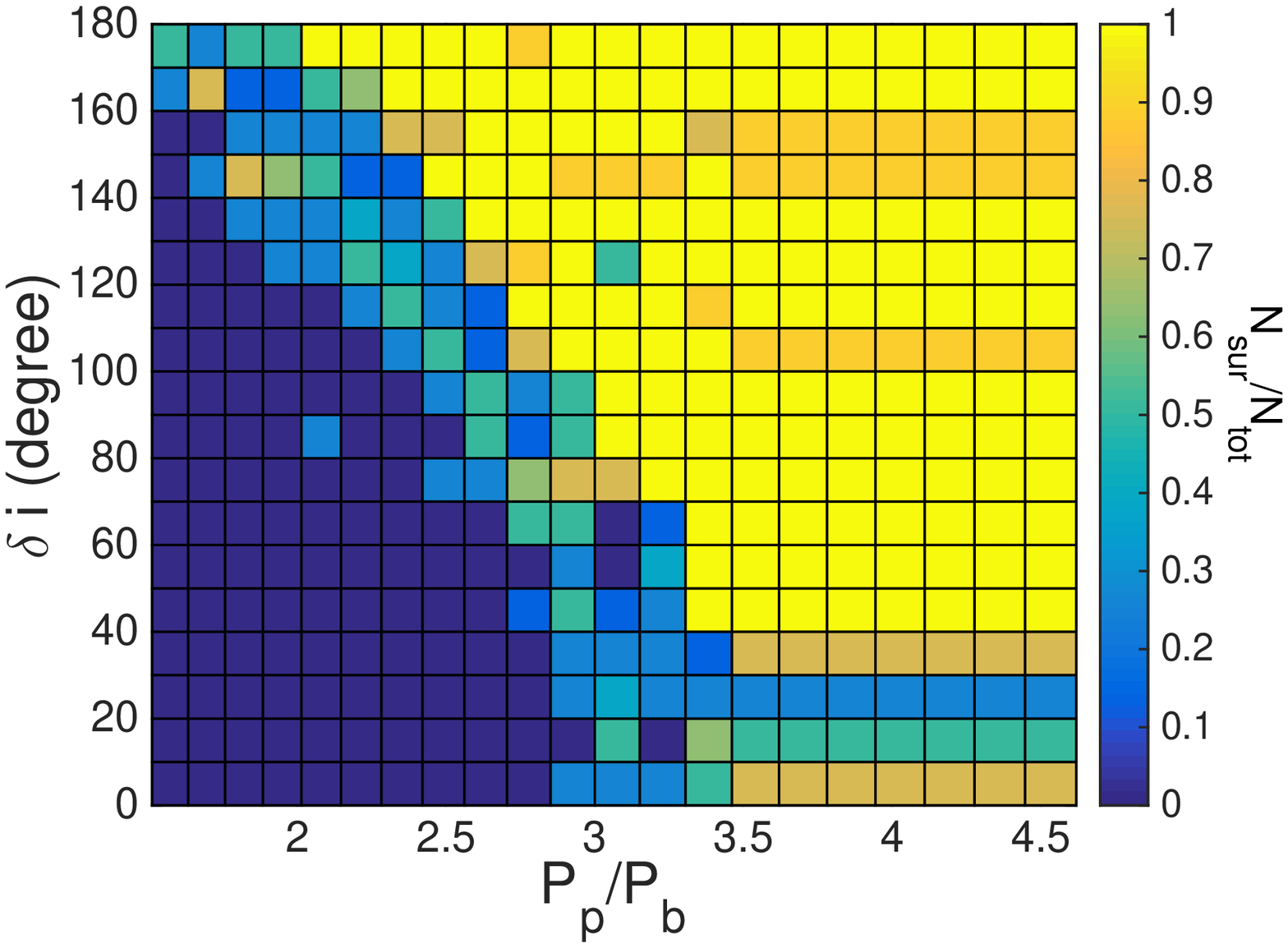} 
\caption{Probability that a system can survive (upper panel) and $\delta a < 0.1 a_0$ (lower panel) as a function of planetary orbital period to stellar orbital period ratio and mutual inclination. The stellar orbital period is 5 days, the mass of the stars are one solar masses and the mass of the planet is 0.001 solar mass. Systems are more stable when the mutual inclination is higher.\label{fig:sur_ai}}
\end{center}
\end{figure}

To study the stability limit when the planetary orbit is misaligned, we performed a large number of numerical integrations. For simplicity, we set the stellar masses to be one solar mass, the stellar orbit to be circular and the planet mass to be a Jupiter mass, and we include simulations with different planetary semi-major axes and the mutual inclinations. For each planetary semi-major axis and mutual inclination, we run 8 simulations with different planetary orbital phase angles equally spaced by $\pi/4$, and we stop the runs after $10^5$ stellar binary periods. We use a 4th-order symplectic integrator (see e.g., \cite{forest1989canonical,suzuki1990fractal, yoshida1990construction, mclachlan2002splitting, tao2016variational}) to obtain the trajectory of the stars and the planets, and we check that the energy change fraction is sufficiently small ($\lesssim10^{-8}$). 

Figure \ref{fig:sur_ai} shows the results of the simulations. The upper panel shows the fraction of survived (stable) systems as a function of initial planetary semi-major axis and mutual inclination. We record that a system is survived (or is stable) when there's no collision and $a_p$ remains within 3 AU. The lower panel shows the probability that the change in planetary semi-major axis is less than $10\%$ of its initial value ($\delta a < 0.1 a_0$) to illustrate the marginally stable region. We find that at higher mutual inclinations, the planetary systems are more stable overall. This is consistent with the results by \citet{Wiegert97}, who studied the stability of planets in alpha Centauri, and with \citet{Doolin11}. Note that instability islands due to resonances can occur at high mutual inclination, when the semi-major axis of the planet still remain outside of the coplanar stability limit, and moderate semi-major axis variations exhibit interesting phase space dependence. Further analysis on these topics is important but is outside of the scope of this article. Since misaligned orbits are also stable inside the stability limit, we use the stability limit as defined in equation (\ref{eqn:sta}) derived by \citet{Holman99} for the coplanar case in the following sections.

\section{Transit Probability}
\label{s:tprob}
Understanding transit probability is important for correcting selection effects in order to obtain the architectural properties of circumbinary planetary systems. Since the observed circumbinary systems so far only involve eclipsing stellar binaries, we focus on the eclipsing stellar binaries in this article. In this section, we first derive the analytical expression of the transit probability for a finite observation period and taking into account orbital precession. Then, we check the analytical expression with numerical simulations.

\subsection{Analytical Expression}
\label{s:ana}
\begin{figure}[h]
\begin{center}
\includegraphics[width=3.3in, height=2.5in]{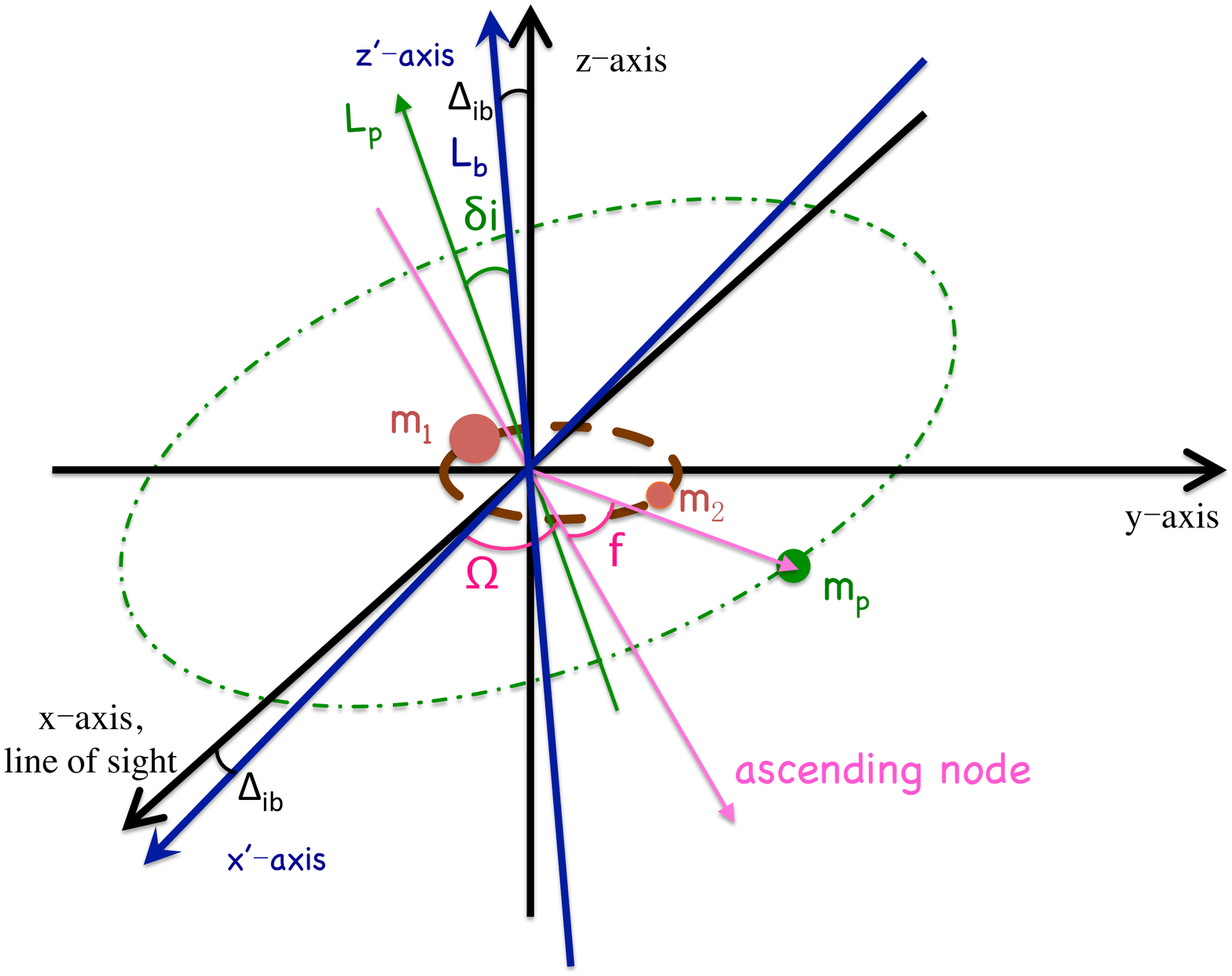} \\
\caption{The configuration of the circumbinary system. $m_1$ and $m_2$ stand for the stellar binary and $m_p$ stands for the planet. $\delta i$ represents the mutual inclination between the orbits of the planet and the stellar binary. \label{fig:config}}
\end{center}
\end{figure}

The configuration of the system is shown in figure \ref{fig:config}, where we align the x-axis with the line of sight, and set the z-axis to be in the plane of the angular momentum of the stellar binary and the line of sight. The axis of $z'$ is aligned with the angular momentum of the stellar binary, and the stellar orbit lies in the plane of $x'-y$. In other words, rotating the $x$ axis and the $z$ axis along the y-axis by the angle $\Delta_{ib} = 90^\circ-i_b$, one can obtain the $x'$ axis and the $z'$ axis separately, where $i_b$ is the line of sight inclination of the stellar orbit. $m_1$ and $m_2$ stand for the masses of the stars and $m_p$ stands for the mass of the planet. $\Omega$ denotes the longitude of ascending node of the planet with respect to the $x'-y$ plane, $f$ denotes the true anomaly of the planet, and $\delta i$ denotes the mutual inclination between the planetary orbit and the stellar binary orbit. Note that we only focus on eclipsing stellar binary, and thus, $\Delta_{ib}$ is very small. In the limit of a circular planetary orbit with semi-major axis ($a_p$), the $x$, $y$, $z$ component of the coordinate of the planet can be expressed as the following:
\begin{align}
x_p &=  a_p(\sin{(f)}\sin{(\delta i)}\sin{(\Delta_{ib})} \\ \nonumber
		& ~~~ +(\cos{(\Omega)}\cos{(f)}-\sin{(\Omega)}\sin{(f)}\cos{(\delta i)})\cos{(\Delta_{ib})}) \\
y_p &= a_p(\sin{(\Omega)}\cos{(f)} + \cos{(\Omega)}\sin{(f)}\cos{(\delta i)}) \\
z_p &= a_p(\sin{(f)}\sin{(\delta i)}\cos{(\Delta_{ib})} \\ \nonumber
	& ~~~ -(\cos{(\Omega)}\cos{(f)}-\sin{(\Omega)}\sin{(f)}\cos{(\delta i)})\sin{(\Delta_{ib})})
\end{align}

With precession, it is difficult to characterize directly the parameter space which allows transits. Thus, we separate the transit criterion into two parts. Specifically, we first use the geometrical approach to obtain the criterion for the planets to transit the stellar binary orbit. Then, we use a probabilistic approach to estimate how often the planet transits the star. This differs from \citet{Martin15}, who considered only the first part, deriving an analytic criterion for orbital crossings, and showed numerically that this guaranteed transits but only with infinite observing time.

To cross the stellar orbit, the line of sight projection of the planet is required to lie within that of the stellar orbit. Specifically, the $y$ component of the planet position needs to be smaller than the semi-major axis of the star ($|y_p| < a_{b, 1}$), the $x$ component of the planet position needs to be positive, and in the limit when the stellar binary is eclipsing ($\Delta_{ib}$ is small), the $z$ component of the planet position needs to be smaller than the radius of the star ($|z_p| < R_{*, 1}$). In the first order in $\Delta_{ib}$ and $\delta i$ the condition to cross the orbit of  $m_1$ is expressed below:
\begin{align}
x_p &\sim a_p(\cos{(\Omega + f)}) > 0, \label{eqn:fmax1} \\
|y_p|  &\sim a_p |\sin(\Omega + f)| < a_{b, 1} , \label{eqn:fmax2} \\
|z_p| &\sim a_p|-\cos{(\Omega + f)} \Delta_{ib} + \sin{(f)}\delta i | < R_{*, 1} \label{eqn:fmax3} .
\end{align}
where $a_{b, 1} = a_{b} m_2/(m_1+m_2)$, $a_b$ is the semi-major axis of the stellar binary orbit, $a_p$ is the semi-major axis of the planetary orbit, and $R_{*, 1}$ is the radius of star $1$. We assume the orbits are circular for simplicity. The expression is interchangeable for $m_2$. In the first order in $a_b/a_p$, \ref{eqn:fmax1}-\ref{eqn:fmax3} can be expressed as:
\begin{align}
-a_{b, 1}/a_p &< \sin(\Omega + f) < a_{b, 1}/a_p , \label{eqn:fmax21} \\
\frac{-R_{*, 1}/a_p + \Delta_{ib}}{\delta i} &< \sin{f} < \frac{R_{*, 1}/a_p + \Delta_{ib}}{\delta i} \label{eqn:fmax22} .
\end{align}
Note that if $-R_{*, 1}/a_p + |\Delta_{ib}| > \delta i$, the planet cannot transit the star $m_1$.

Extending to higher $\delta i$ by substituting $\delta i$ with $\sin{(\delta i)}$ in equation \ref{eqn:fmax21}-\ref{eqn:fmax22}, we obtain the range of $\Omega$ that the planet can cross the stellar orbit of $m_1$:
\begin{equation}
\Delta \Omega_{1} = 
	\begin{cases}
			\min{[2(f_2-f_1)+4{\rm asin}(a_{b, 1}/a_p), 2\pi]} \\ 
			~~~\text{if $\pi/2-f_2>{\rm asin}(a_{b, 1}/a_p)$} \\
			~~~~~\text{\& $f_1+\pi/2>{\rm asin}(a_{b, 1}/a_p)$}, \\
			\min{[2(f_2-f_1)+2(\pi/2-f_2)+2{\rm asin}(a_{b, 1}/a_p), 2\pi]} \\  
			~~~\text{else if $f_1+\pi/2>{\rm asin}(a_{b, 1}/a_p)$}, \\
			\min{[2(f_2-f_1)+2(\pi/2-f_2)+2(f_1+\pi/2), 2\pi]} \\  
			~~~\text{else} .
	\end{cases}
\end{equation}
where 
\begin{align}
f_1 = \begin{cases}
			{\rm asin}\Big(\frac{-R_{*, 1}/a_p+\sin(|\Delta_{ib}|)}{\sin(\delta i)}\Big) \\
			~~~\text{if $-1<(-R_{*, 1}/a_p+\sin(|\Delta_{ib}|))/\sin(\delta i)<1$} ,\\
			-\pi/2  \\
			~~~\text{else if $(-R_{*, 1}/a_p+\sin(|\Delta_{ib}|)/\sin(\delta i)<-1$} ,\\  
			\pi/2    \\
			~~~\text{else} .
	\end{cases}.
\end{align}
and 
\begin{align}
f_2 = \begin{cases}
			{\rm asin}\Big(\frac{R_{*, 1}/a_p+\sin(|\Delta_{ib}|)}{\sin(\delta i)}\Big) \\
			~~~\text{if $-1<(R_{*, 1}/a_p+\sin(|\Delta_{ib}|))/\sin(\delta i)<1$} ,\\
			-\pi/2  \\
			~~~\text{else if $(R_{*, 1}/a_p+\sin(|\Delta_{ib}|)/\sin(\delta i)<-1$} ,\\  
			\pi/2    \\
			~~~\text{else} .
	\end{cases}.
\end{align}
The difference of $f_2$ and $f_1$ contributes to the range of $\Omega$ that allows the planet to cross the stellar orbit. Note that $\Omega$ has two distinctive regions which allow transit if $\pi/2-f_2>{\rm asin}(a_{b, 1}/a_p)$, and these two regions are connected if $f_1+\pi/2>{\rm asin}(a_{b, 1}/a_p)$. 

Next, we take into account orbital precession to estimate the probability that the planet will cross the stellar orbit over time. Briefly, orbital precession can increase the transit probability, because it broadens the range of $\Omega$ covered, increasing the likelihood of it entering the window which allows transits. Since most of the stellar binaries are not highly eccentric, we take the limit when $e_b \to 0$. Then, the precession timescale ($T_{prec}$) of the planetary orbit scales with the planetary orbital period ($P_p$) as:
\begin{align} 
T_{prec} &= \frac{2\pi}{|\dot{\Omega}|} \\
&=  P_p \frac{4}{3 \cos{\delta i}}\frac{a_p^2}{a_b^2}\frac{(m_1+m_2)^2}{m_1m_2} , \label{eqn:prec}
\end{align}
as obtained by \citet{Schneider94}, who considered the precession timescale when $m_1 = m_2$. \citet{Farago09} derived an equation for the more general case with eccentric binaries. Note that when the stellar binary is eccentric, the precession is more complicated, as the longitude of node can librate around $\pm\pi/2$ when the inclination is high. When the stellar binary is circular, $\Omega$ decreases with time when the inclination is below $\pi/2$ and $\Omega$ increases with time when the inclination is above $\pi/2$. We adopt the expression of equation (\ref{eqn:prec}) for simplicity in the following analysis, and $\Omega$ increases/decreases linearly with time. Specifically, the change in $\Omega$ due to precession is denoted as $\delta \Omega_{prec} = \dot{\Omega} T_{obs}$.

The total range of the longitude of node during observation time period ($T_{obs}$) is 
$\delta \Omega_{1} = \Delta \Omega_1 + \dot{\Omega} T_{obs}$. 
\begin{equation}
\delta \Omega_1 = \begin{cases}
   \Delta \Omega_{1} + 2\delta \Omega_{prec} \\ 
   ~~~\text{if $\pi/2-f_2>\delta \Omega_{prec}/2$} \\ 
   \Delta \Omega_{1} +\delta \Omega_{prec}+2(\pi/2-f_2) \\ 
    ~~~\text{else if $\pi/2-f_2>{\rm asin}(a_{b, 1}/a_p)$} \\
    \Delta \Omega_{1} +\delta \Omega_{prec} \\ 
    ~~~\text{else}
\end{cases}
\end{equation}

Then, the probability to cross the stellar orbit ($P_{cr, 1}$) for $m_1$ is:
\begin{equation}
P_{cr, 1} =  \frac{\min{[\delta \Omega_1, 2\pi]}}{2\pi} .
\end{equation}

We next calculate the probability for the planet to transit the stars given that the planet crosses the stellar orbit. It is roughly the ratio of the relative displacement of the planet and the star as the planet crosses the orbit to the projected width of the stellar orbit. The relative displacement depends on the time it takes for the planet to cross the orbit, which can be expressed as $t_{trans} \sim \pi/2(R_{*, 1}) / (v_p \sin{\delta i})$, where $v_p$ is the orbital velocity of the planet, and the factor $\pi/2$ corresponds to the correction taking into account the different impact parameters to cross the star. The relative velocity of the planet and the star depends on whether they are on the same side of the star. When the planet and the star are both towards the observer with respect to the center of mass ($x_p > x_* > 0$), the planet and the star are moving in the same direction, and the relative displacement is roughly: $dl_1 = \big( t_{trans} (|v_p \cos{\delta i} - 2 v_{*, 1} / \pi |) + R_{*, 1} \big) $, where $v_{*, 1}$ is the orbital velocity of the star, and the factor $2/\pi$ gives the averaged line of sight projected stellar velocity. On the other hand, when the star is on the other side of the center of mass ($x_p > 0 > x_* $), the planet and the star move in the opposite directions. Thus, the relative displacement is $dl_2 = \big( t_{trans} (|v_p \cos{\delta i} + 2 v_{*, 1} / \pi |) + 2 R_{*, 1} \big)$. The projected size of the stellar orbit can be expressed as: $2a_{b, 1}$. Therefore, the probability that the planet can transit in front of the star $m_1$ is roughly:
\begin{align}
P_{*, 1} &= 
\begin{dcases}
    1 ,& \text{if } (dl_1+dl_2)/2 > 2 a_{b, 1} \\
    \frac{1}{4 a_{b, 1}}(dl_1+dl_2) & \text{otherwise} 
\end{dcases}
\end{align}

Note that the movements of the planet and the star enhance the transit probability. Specifically, the motion of the planet is important when the mutual inclination is low, as it takes a long time for the planet to cross the stellar orbits. In addition, the relative displacement and the transit probability are dominated by the motion of the stars when the mutual inclination is high. 

During the observation time period $T_{obs}$, the planet can cross the stellar orbit multiple times, and the transit probability increases as the number of orbit crossing increases. The maximum number of crossing is $T_{obs}/P_p$, which occurs when the precession is slow and the longitude of node stay in the window that allows crossing (e.g., when the mutual inclination is high $\sim 90^\circ$). In this limit, the planet crosses the stellar orbit every time. When the precession is fast, the precession dominates the number of stellar orbit crossing time, and the number of orbit crossing is $\Delta \Omega_{1}/(\dot{\Omega} P_p)$ if $\pi/2-f_2<{\rm asin}(a_{b, 1}/a_p)$, when the region where $\Omega$ allows transit over $2\pi$ is connected, and $\Delta \Omega_{1}/2/(\dot{\Omega} P_p)$ if $\pi/2-f_2>{\rm asin}(a_{b, 1}/a_p)$, when the regions where $\Omega$ allows transit over $2\pi$ are separated. When the precession is even faster, the node can precess to the range that allows transit more than once. In sum, the number of stellar orbit crossing can be expressed as the following:
\begin{equation}
n_{1} = 
\begin{cases}
    \min[ \frac{T_{obs}}{P_p}, \frac{\Delta \Omega_1/2 + \delta \Omega_{prec}} {\pi} \frac{\Delta \Omega_{1}/2}{\dot{\Omega} P_p}] \\
    ~~~ \text{if $\pi/2-f_2>{\rm asin}(a_{b, 1}/a_p)$} \\
    ~~~\text{\& $\Delta \Omega_1/2 + \delta \Omega_{prec} >\pi$} \\
    \min[ \frac{T_{obs}}{P_p}, \frac{\Delta \Omega_{1}/2}{\dot{\Omega} P_p}] ,                  \\
    ~~~\text{else if $\pi/2-f_2>{\rm asin}(a_{b, 1}/a_p)$} \\
    \min[ \frac{T_{obs}}{P_p}, \frac{\Delta \Omega_1 + \delta \Omega_{prec}}{2\pi} \frac{\Delta \Omega_{1}}{\dot{\Omega} P_p}] \\
    ~~~\text{else if $\Delta \Omega_1 + \delta \Omega_{prec} >2\pi$} \\
    \min[ \frac{T_{obs}}{P_p}, \frac{\Delta \Omega_{1}}{\dot{\Omega} P_p}] \\
    ~~~\text{else.}    
    \end{cases}
\end{equation}
Assuming each orbit crossing is independent, the probability to transit the star $m_1$ at least once is:
\begin{equation}
\label{eqn:tprob1}
P_{cr, 1} (1-(1-P_{*, 1})^{n_{1}}) .
\end{equation}
Equation (\ref{eqn:tprob1}) can be applied to systems involving a faint star, where only the transit of the primary star can be detected.

Considering transits of both stars, one can obtain the probability for the planet to transit the two stars at least once:
\begin{align}
 \label{eqn:tprob}
P_{transit} = P_{cr, 2} \Big[1-(1-P_{*, 2})^{n_{2}} \Big(\frac{P_{cr, 1}}{P_{cr, 2}} (1-P_{*, 1})^{n_{1}}  \\
+ \frac{P_{cr, 2} - P_{cr, 1}} {P_{cr, 2} }\Big)\Big]   \nonumber
\end{align}
where $m_1 > m_2$, since if a planet crosses the orbit of $m_2$, it can also cross the orbit of $m_1$. For simplicity, we assume that if the planet crosses both stellar orbits, the transit events are independent. Note that the independence approximation do not generally hold, since the stars are $180^\circ$ out of phase with each other, and each orbit crossing has a roughly fixed phase difference between each other, due to the periodic nature of the stellar orbits. However, we illustrate in the following section (\textsection \ref{s:num}), that the transit probability obtained using the independence approximation agrees well with the numerical results.

In addition, one can also calculate the average number of transits given that a system transits. The average number of transits can help determine the likelihood to detect the transit events, since it's more likely to detect the planet when the transit number increases. Specifically, the averaged number of transits can be expressed as the following:
\begin{align}
\label{eqn:Nprob}
N_{transit} =& n_{2} P_{*, 2} + n_{1} P_{*, 1} P_{cr, 1} / P_{cr, 2} ,
\end{align}
where similar to equation (\ref{eqn:tprob}), we also assume $m_1 > m_2$. As shown in the following section (\textsection \ref{s:num}), the analytical expression agrees with the numerical results, except at low mutual inclination, where the independence approximation causes the averaged number of transits to be larger than the numerical results. Specifically, the expected number of transits differ within a factor of two (see more discussions in section \textsection \ref{s:num}). 

\subsection{Numerical Comparison}
\label{s:num}
To test how well the analytical expression predicts the transit probability, we compare the analytical results with the numerical transit probabilities obtained from numerical simulations. We include three sets of planetary systems for illustration: planetary systems with equal-mass stellar binaries with line of sight inclination at $90^\circ$(\ref{s:esb}), planetary systems with eclipsing equal-mass stellar binaries ($i_b$ near but not exactly $90^\circ$) (\ref{s:ecl}), and the observed planetary systems with un-equal mass stellar binaries in eccentric orbits around eclipsing binaries (\ref{s:usb}). 

\subsubsection{Equal-mass Stellar Binaries along Line of Sight}
\label{s:esb}
In this section, we consider the transit probability for planetary systems composed of two solar type stars in a circular orbit, surrounded by planets with different semi-major axes and mutual inclinations. The transit probability depends sensitively on the mutual inclination between the stellar orbit and the planetary orbit. We thus first check the analytical results with numerical results including different mutual inclinations. In this section, we set the stellar binary to be aligned with the line of sight first, and we relax this assumption in the next section (\textsection \ref{s:ecl}). Specifically, for each mutual inclination, we run 1000 numerical simulations with planetary true anomaly ($f$) and longitude of ascending node ($\Omega$) randomly drawn from a uniform distribution. We record the number of systems in which the planet transits in front of the stars to obtain the transit probability for each mutual inclination. 

\begin{figure}[h]
\begin{center}
\includegraphics[width=3.3in, height=2.5in]{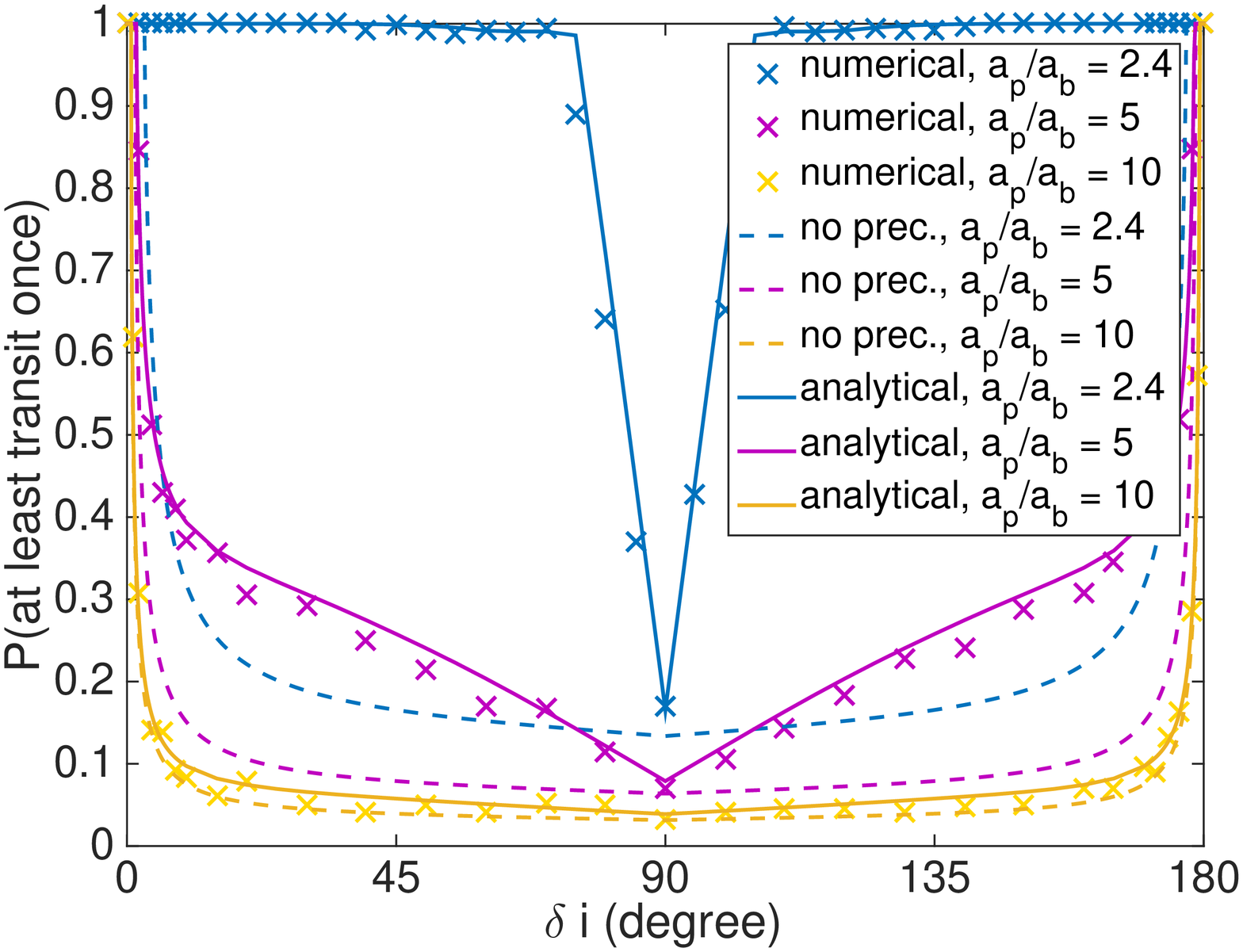} \\
\includegraphics[width=3.3in, height=2.5in]{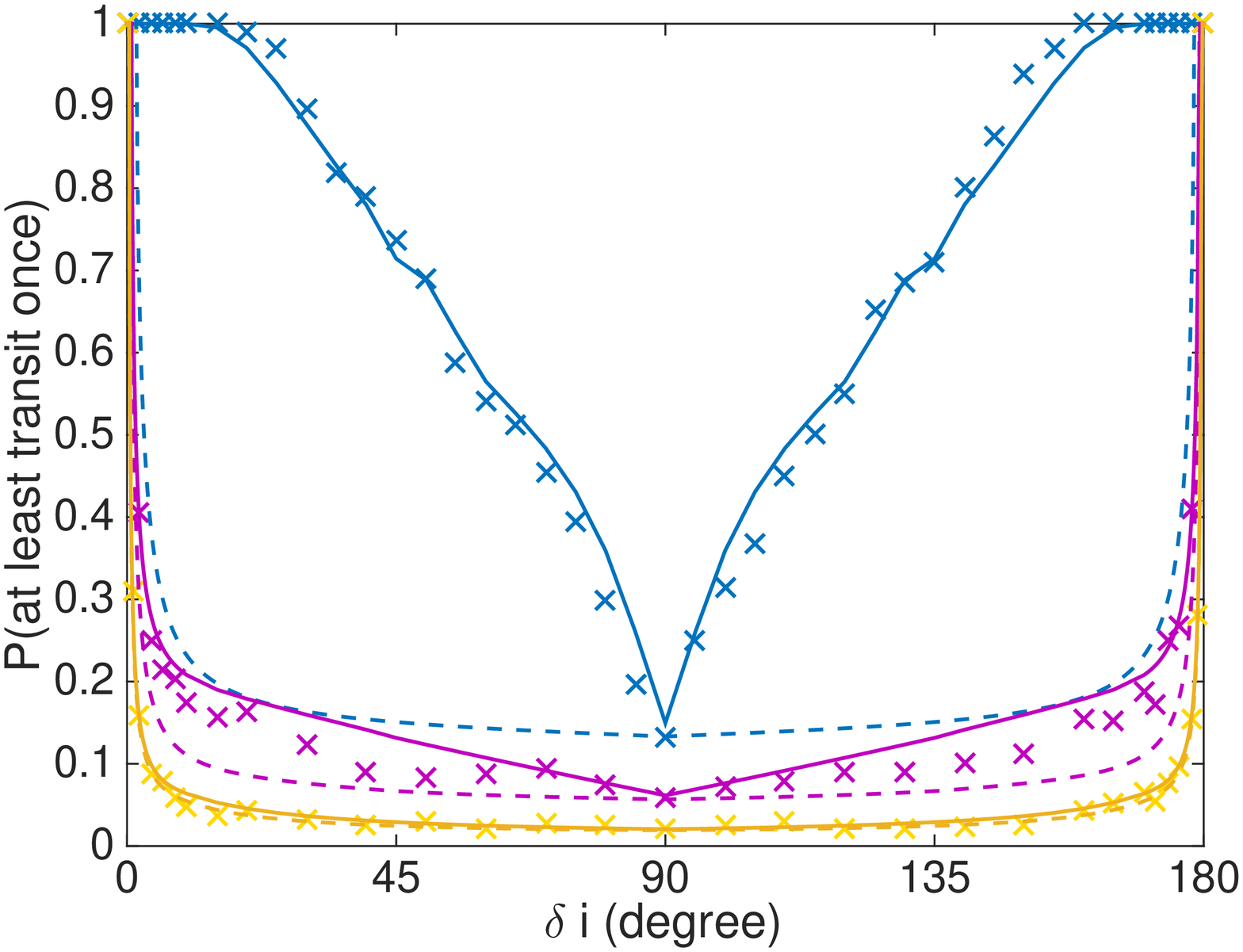} 
\caption{Upper panel: transit probability in one year for circumbinary planets surrounding a $P=2$ day stellar binary as a function of $\delta i$; lower panel: transit probability for planets surrounding a $P=5$ day stellar binary as a function of $\delta i$. The solid lines indicate the analytical results, and the crosses are the numerical results. The dashed lines represent the case without precession. The analytical results agree well with the numerical simulation, and the transit probability is greatly under-predicted without precession.  \label{fig:Pi}}
\end{center}
\end{figure}

The upper panel of figure \ref{fig:Pi} shows the probability to transit either of the binary stars at least once in one year when the orbital period of the stellar binaries is two days, and the lower panel shows the case when the orbital period is five days. The solid lines represent the results of the analytical expression we derived in \textsection \ref{s:ana} (see equation (\ref{eqn:tprob})), and the crosses are the numerical results. The different colors represent the different planet to stellar semi-major axis ratios. The blue crosses represent the probability when the planetary semi-major axis is $2.4$ times that of the stellar binary, where the critical semi-major axis for stability is $\sim 2.4 a_b$ in these cases, according to the stability limit by equation (\ref{eqn:sta}) \citep{Holman99}. The purple crosses represent the case when $a_p / a_b = 5$, and the yellow crosses represent the case when $a_p / a_b = 10$. For all the cases included here, the analytical results agree very well with the numerical results. In addition, different from the case with infinite amount of observation time, where \citet{Martin15} found that the transit probability increases as the mutual inclination increases, the transit probability decreases as the mutual inclination increases here in the finite observation time case when $i_b = 90^\circ$. 

The dashed lines in figure \ref{fig:Pi} represent the case when we ignore orbital precession. At low mutual inclinations, the precession timescale is shorter and the parameter space that allows transit increases. Therefore, at lower mutual inclinations, the transit probability is much higher when one includes orbital precession. This agrees with \citet{Martin15}. At $\sim 90^\circ$, the precession time is long, and the results with and without orbit precession become similar. Moreover, the precession timescale increases steeply with the planetary orbital period. Thus, when the semi-major axis of the planet is larger, the difference between the case with and without precession also becomes smaller.

\begin{figure}[h]
\begin{center}
\includegraphics[width=3.3in, height=2.5in]{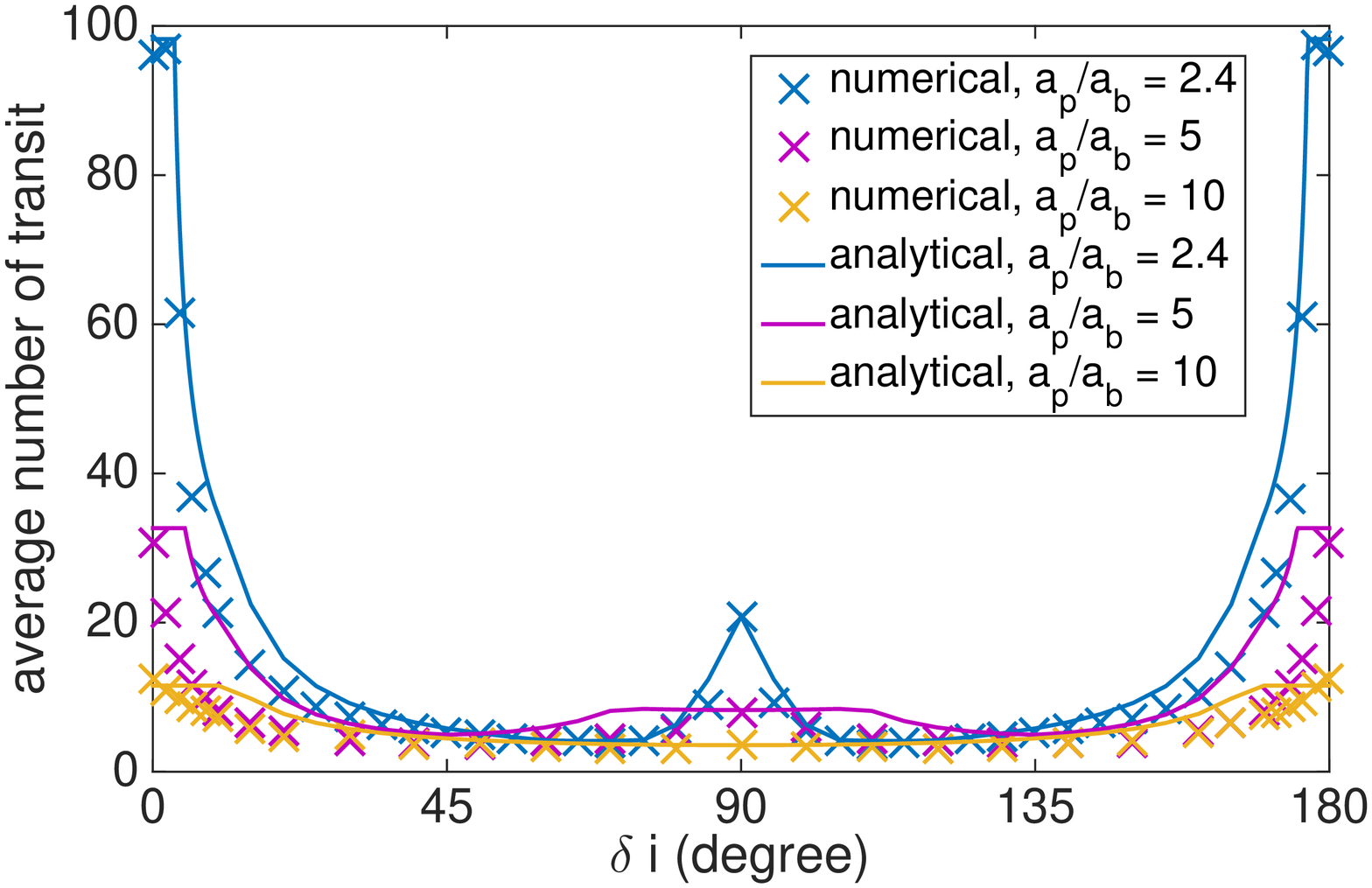} \\
\includegraphics[width=3.3in, height=2.5in]{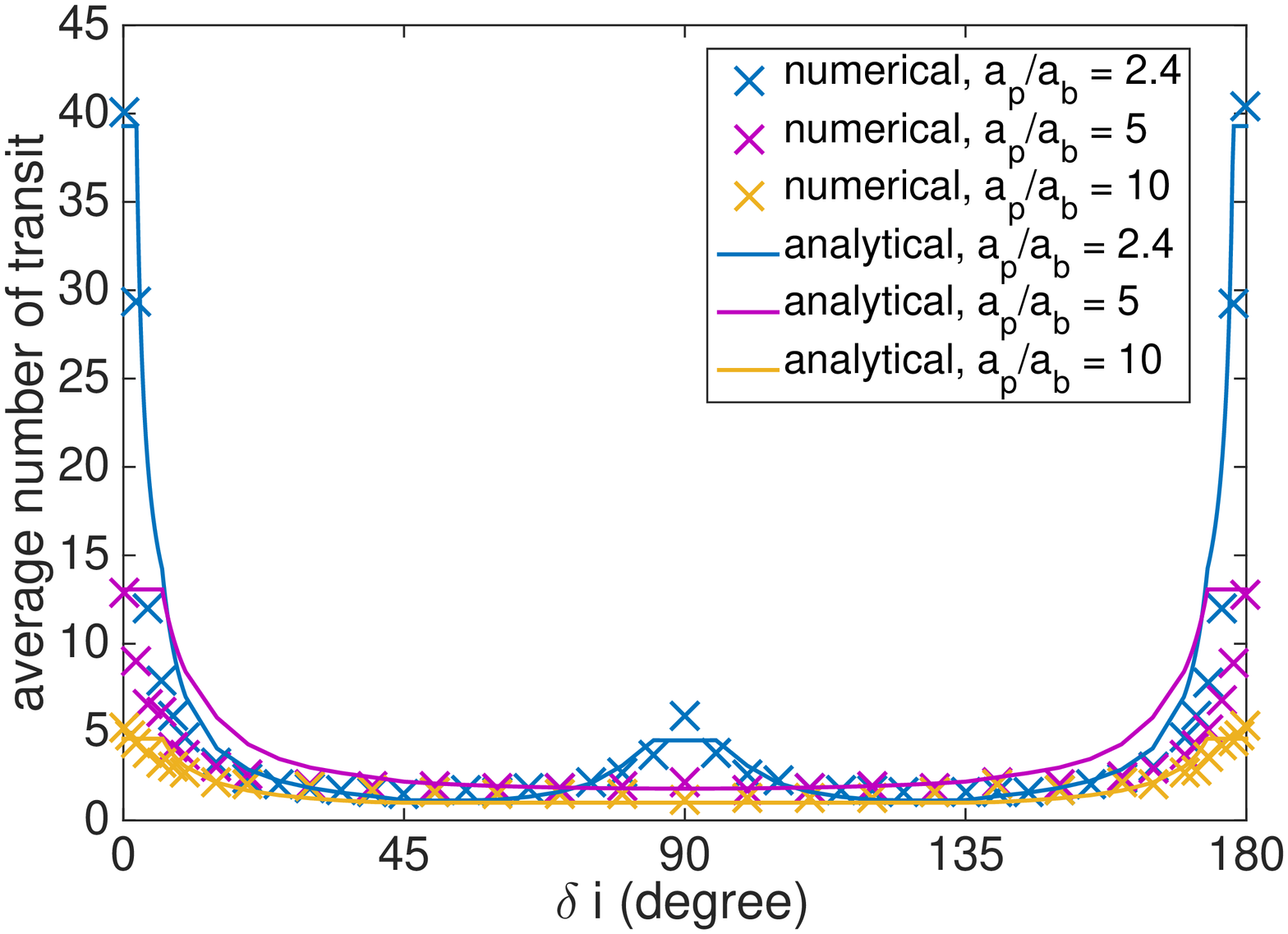} 
\caption{Upper panel: average number of transits given the system transits at least once surrounding a $P=2$ day stellar binary as a function of $\delta i$; lower panel: average number of transit given the system transits at least once surrounding a $P=5$ day stellar binary as a function of $\delta i$. The analytical expression over-estimates the number of transit within a factor of two at low mutual inclinations. \label{fig:MeanN}}
\end{center}
\end{figure}

The number of transits is important for planet detection, since the more the planet transits the stars, the more likely it can be detected. For instance, the detected circumbinary systems all have at least three primary and/or secondary transits upon publication. To investigate this, we record the number of transits for each system in the simulation to obtain the average number of transits for each mutual inclination given that the planet transits at least once. The results are shown in Figure \ref{fig:MeanN}. Similar to figure \ref{fig:Pi}, the upper panel shows the case when the planet orbits a circular two-day stellar binary with solar masses and solar radii, and the lower panel shows the case when the planet orbits a circular five-day stellar binary. We also include planets with three different semi-major axes ($a_p = 2.4 a_b$, $a_p = 5 a_b$ and $a_p = 10 a_b$) represented by different colors. As expected, the average number of transits is smaller when the planetary orbit is farther from the stellar binary, where the range of longitude of node that allows transits is smaller and the orbital period is longer. Interestingly, the average number of transits is larger when the mutual inclination is around $\sim 90^\circ$. This is because the precession time is long when the mutual inclination is higher, and the longitude of node will stay longer in the range that allows transits, which leads to a higher average number of transits for planetary systems that transit at least once. 

Comparing the analytical results (solid lines, equation (\ref{eqn:Nprob})) with the numerical results (crosses), figure \ref{fig:MeanN} shows that at low mutual inclinations, the analytical expression systematically leads to a larger average number of transits than the numerical results. This is because the transit events of the two stars are not independent, as assumed in the analytical derivation. Since the two stars are $180^\circ$ out of phase with each other, the likelihood for both of the stars to transit is reduced. This reduction is important when the mutual inclination is lower, where $P_{*,1~or~2}$ is so large that the likelihood for the two stars to both transit is high assuming the transits are independent. Since the reduction may change a double transits to a single transit during one planetary orbit period, it can at most decrease the average number by a factor of two. Thus, as shown in figure \ref{fig:MeanN}, the average number of transits from the analytical expression is consistent within a factor of two from the numerical results.

\begin{figure*}[h]
\begin{center}
\includegraphics[width=6.6in, height=3.in]{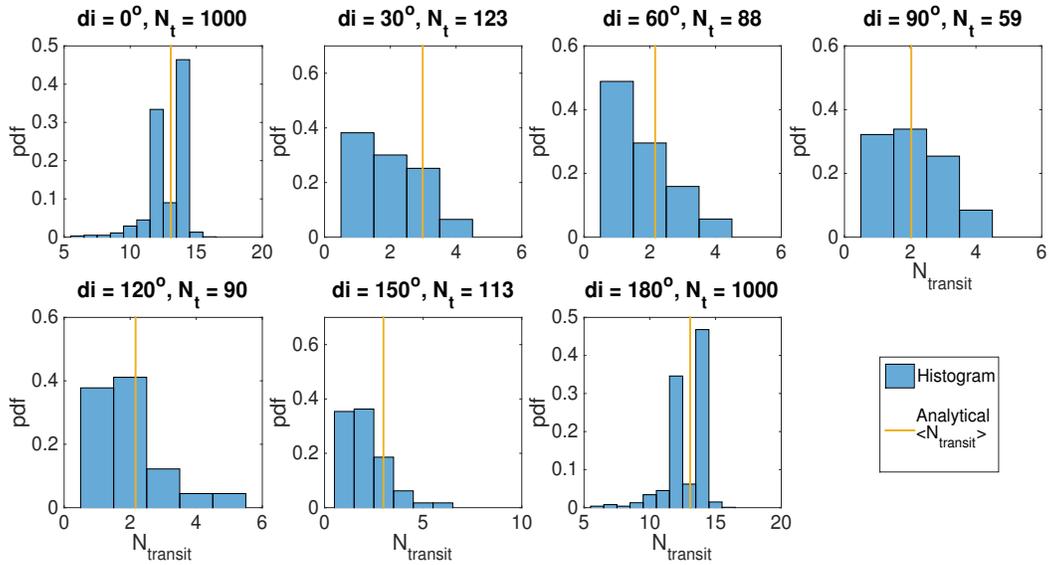} 
\caption{Distribution of the number of transits in one year. The planetary semi-major axis is set to be five times that of the stellar binary ($a_p/a_b = 5$) and the stellar binary period is five days $P_* = 5$ days. Each panel corresponds to a different mutual inclination, and the number of systems that transit at least once are shown in the title of each panel. \label{fig:Ntrans}}
\end{center}
\end{figure*}

To illustrate the distribution of the number of transits, figure \ref{fig:Ntrans} shows the histogram of the number of transits for the circumbinary planetary systems with a five-day stellar orbit, and a planet to stellar semi-major axis ratio of five. Each panel corresponds to a different mutual inclination ($\delta i$), and we include 1000 systems with random orbital phases to obtain the results in the histogram. The widths of the bins are unity, and the number of systems that transit at least once ($N_t$) is shown in the title of each panel. The yellow solid line represent the average number according to the analytical expression (see equation (\ref{eqn:Nprob})).

The results in figure \ref{fig:Ntrans} can be understood from geometrical interpretations. Specifically for this case, the planet period is roughly $56$ days, and in one year, the planet orbits $6.5$ times. Thus, the maximum number of transits is $2 \times (6+2) = 16$ times, where the planet transits both stars before and after the full orbits, agreeing with the numerical results shown in figure \ref{fig:Ntrans}. When the mutual inclination is low $\sim 0^\circ$, the planet transits the stars every orbit. Since the stars may overlap in projection during the transit, the minimum number of transits is $6$ times. Moreover, the average number of transits peaks around $2 \times 6 = 12$ and $2 \times 7 = 14$ times when the mutual inclination is 0 or $180^\circ$, as shown in figure \ref{fig:Ntrans}. At high mutual inclinations, the histograms show that the planet still have a high probability to transit at least twice, and thus it is unlikely to miss the transits. The average number of transits is symmetric with respect to $90^\circ$, and the deviation from this symmetry is due to the random fluctuations. 

\subsubsection{Eclipsing Equal-mass Stellar Binaries}
\label{s:ecl}
Since the stellar binaries do not need to be aligned at $i_b = 90^\circ$ to be eclipsing, we also consider the case when $i_b$ near but not exactly $90^\circ$. This is very different from the case when $i_b = 90^\circ$, especially for low $\delta i$. Specifically, the planet cannot transit the stars when the mutual inclination $\delta i <  i_{p, c, 1} = |\Delta i_b| - {\rm asin}{[(a_{b, 1}\sin{{\Delta i_b}}+R_{*, 1})/a_p]}$, where $\Delta i_b = 90-i_b^\circ$, as mentioned in \textsection \ref{s:ana}. Considering the transit of both stars, $P_{transit} = 0$ when $\delta i < \min[i_{p, c, 1}, i_{p, c, 2}]$.

\begin{figure}[h]
\begin{center}
\includegraphics[width=3.6in]{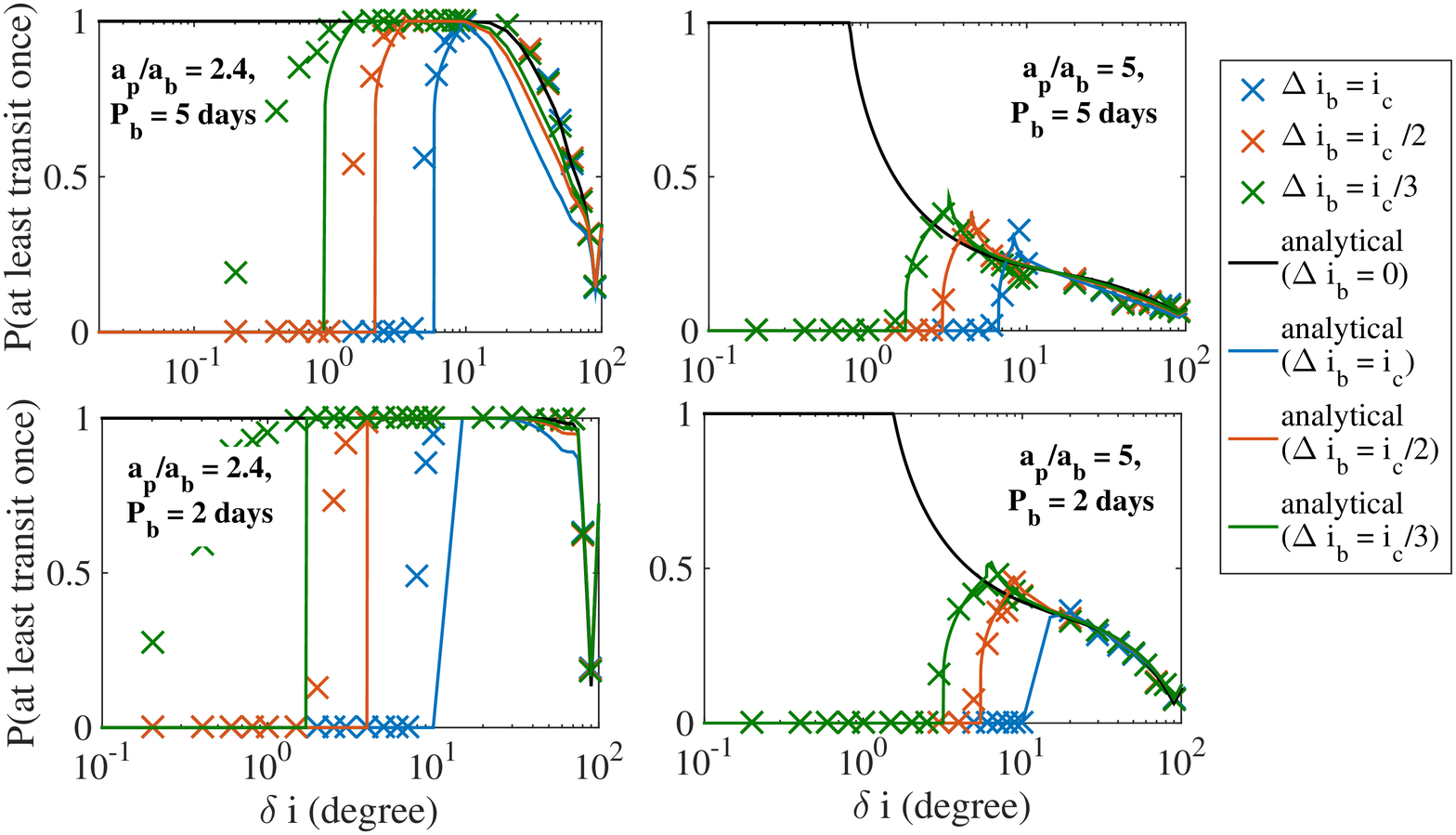} 
\caption{Probability that a planet transits at least once in one year as a function of $\delta i$ when $a_p = 2.4 a_b$ (left panels) and when $a_p = 5 a_b$ (right panel). The crosses represent the numerical results and the solid lines represent the analytical results. The analytical expression do not fit well when $a_p = 2.4 a_b$, because the planetary orbit is no longer circular due to the strong perturbation of the stellar binary.  \label{fig:Rotate}}
\end{center}
\end{figure}

To compare the analytical expression of the transit probability with numerical results, we set the binary star to be sun-like, in a circular orbit and with a orbital period of two days and five days, and we simulate the case when the planet semi-major axis is set to be $a_p = 2.4 a_b$ and $a_p = 5 a_b$, similar to \textsection \ref{s:esb}. The critical line of sight inclination of the stellar binary to be eclipsing is $i_c = 2R_{\odot}/a_b = 13.7^\circ$ when the stellar binary is in a 5-day orbit, and $i_c = 2R_{\odot}/a_b = 7.4^\circ$ when the stellar binary is in a 5-day orbit. To include different stellar inclination, we set $i_b = i_c$, $i_b = i_c/2$ and $i_b = i_c/3$ separately. The results are shown in figure \ref{fig:Rotate}.
 
Figure \ref{fig:Rotate} shows the transit probability over one year. The numerical results are indicated by crosses and the analytical results are represented by the solid lines. Note that when $a_p/a_b = 2.4$, the planet can still transit when $\delta i$ is smaller than the critical value. This is because the minimum separation $r_{p, min}$ is smaller than the semi-major axis of the planet, $a_p$, since the planetary orbit is not circular and the semi-major axis can vary from its initial value due to the perturbation of the stellar binary. Thus, the analytical results do not agree well with the numerical results when $a_p/a_b = 2.4$. The discrepancy is smaller when $|\Delta_{ib}|$ is smaller. In addition, figure \ref{fig:Rotate} shows that when $\delta i$ is large, the transit probability approaches the results when $i_b$ is set to be $90^\circ$. 

It was found by \citet{Martin15} that taking an infinite amount of time, the transit probability of circumbinary planets is higher than that of planets orbiting a single star. However, this is not always true for a finite observation time, especially when the total observation time is very short. For instance, the precession of the planetary orbit is faster for the circumbinary planets and allows a larger parameter space for the planet to cross the stellar orbit, yet the planet may not transit the star when it crosses the binary orbit. Specifically, the probability to transit a single star is $P = R_* / a_p$ for a circular planetary orbit \citep{Borucki84}. For planets around a 5-day orbital period stellar binary at $a_p/a_b = 2.4$, the one month transit probability of the circumbinary planets is higher than the case if we substitute the stellar binary by a single solar-type star, where the probability to transit is $P = R_\odot / a_p = 0.027$. However, at $a_p/a_b = 5$, the transit probability of the circumbinary planets in one month is lower than that orbiting a single solar-type star when the mutual inclination between the planet and the binary star is high ($\gtrsim 50^\circ$). When the total observation time is increased to $\gtrsim 60$ days, the probability to transit is lower for the single star case, even at $a_p/a_b = 5$ for all $\delta i$. This suggests that for the {\it TESS} mission, some of the circumbinary planets may have lower transit probabilities than those of their counterpart planets around single stars.

\subsubsection{Observed Circumbinary Planets}
\label{s:usb}
To test the analytical expression when the stellar binaries are composed of stars with different stellar masses and in eccentric stellar orbits, we obtain the transit probability for the observed circumbinary planets numerically and compare the analytical results with the numerical simulation. We allow the stellar inclination $i_b \neq 90^\circ$ in this section. The properties of the circumbinary planets are listed in Table \ref{t:cir}. Most of the planetary orbits are nearly circular, except Kepler-47c. However, the eccentricity of Kepler-47c is quite uncertain. Thus, for simplicity, we set the planetary orbits to be circular.

\begin{table*}[h]
\caption{Properties of observed transiting cimcumbinary planets. \label{t:cir} }
\centering
\begin{tabular}{ |l|l|l|l|l|l|l|l|l|l|l|l|l| }
    \hline
       & $m_1 (M_{\odot})$ & $m_2 (M_{\odot})$  & $R_{*,1} (R_{\odot})$ & $R_{*,2} (R_{\odot})$ & $a_b (AU)$ & $e_b$ & $m_p (M_{\rm J})$ & $R_p (R_{\rm J})$ & $a_p (AU)$  & $\delta i (^\circ)$ & $e_p$ & $i_b (^\circ)$\\ \hline
Kepler 16b\footnote{data obtained from Table 1 of \citet{Doyle11}. $a_p$ differs from \citet{Martin15}, who set $a_p = 0.71$ AU. $\delta i$ is obtained from Table 1 of \citet{Martin15}.} 
& 0.69 & 0.20 & 0.65 & 0.23 & 0.22 & 0.16 & 0.33 & 0.75 & 0.70 & 0.31 & 0.0069 & 90.34\\ \hline
Kepler 34b\footnote{data obtained from Table 1 of \citet{Welsh12}. $R_{*,2}$ differs from \citet{Martin15}, who had a typo and set $R_{*,2} = 0.19 R_{\odot}$ in their Table 1. $\delta i$ is obtained from table 1 of \citet{Martin15}.}
& 1.05 & 1.02 & 1.16 & 1.09 & 0.23 & 0.52 & 0.22 & 0.76 & 1.09 & 1.86 & 0.18 & 89.86 \\ \hline
Kepler 35b\footnote{data obtained from Table 1 of \citet{Welsh12}. $\delta i$ is obtained from table 1 of \citet{Martin15}. }
& 0.89 & 0.81 & 1.03 & 0.79 & 0.18 & 0.14 & 0.13 & 0.73 & 0.60 & 1.07 & 0.042 & 90.42 \\ \hline
Kepler 38b\footnote{data obtained from Table 6 of \citet{Orosz12b}. $m_2$ and $R_{*,1}$ differ from \citet{Martin15}, who set them to be $0.27 M_{\odot}$ and $1.78 R_{\odot}$ separately. $m_p$ is set to be 0.38 $M_{\rm J}$ in the numerical simulation. The results are not sensitive to the planet mass because $m_p \ll m_{1, 2}$} 
& 0.95 & 0.25 & 1.76 & 0.27 & 0.15 & 0.10 & $<0.38$ & 0.39 & 0.46 & 0.18 & $<0.032$ & 89.27 \\ \hline
Kepler 47b\footnote{data obtained from table 1 and from the main text of \citet{Orosz12a}. We adopt $m_p = 0.031M_{\rm J}$ for numerical simulation. The results are not sensitive to the planet mass because $m_p \ll m_{1, 2}$. $m_2$, $R_{*,1}$ and $R_{*,2}$ differ from \citet{Martin15}, who set $m_2$ to be $0.46M_{\odot}$, $R_{*,1}$ to be 0.84 $R_{\odot}$ and $R_{*,2}$ to be 0.36 $R_{\odot}$.} 
& 1.04 & 0.36 & 0.96 & 0.35 & 0.084 & 0.023 & $0.022-0.031$ & 0.27 & 0.30 & 0.27 & $<0.035$ & 89.34 \\ \hline
Kepler 47c\footnote{same as K-47b, we obtain data from \citet{Orosz12a}, we set $m_p = 0.072M_{\rm J}$ for numerical simulation. The results are not sensitive to the planet mass because $m_p \ll m_{1, 2}$} 
& 1.04 & 0.36 & 0.96 & 0.35 & 0.084 & 0.023 & $0.050-0.072$ & 0.42 & 0.99 & 1.16 & $<0.41$ & 89.34 \\ \hline
Kepler 64b\footnote{\citet{Schwamb13} and \citet{Kostov13}. The results of both studies are consistent with each other. For the simulations, we use the results of \citet{Schwamb13}, and we set $m_p = 0.531 M_{\rm J}$ in the numerical simulation. $m_{*,1}$, $m_{*,2}$, $R_{*,1}$, $R_{*,2}$, $a_b$ and $a_p$ all differ from \citet{Martin15}, who set $m_{*,1}=1.50$, $m_{*,2}=0.40$, $R_{*,1}=1.75$, $R_{*,2}=0.42$, $a_b=0.18$ and $a_p=0.65$} 
& 1.53 & 0.41 & 1.73 & 0.38 & 0.17 & 0.21 & $<0.531$ & 0.56 & 0.63 & 2.81 & 0.054 & 87.36 \\ \hline
Kepler 413b\footnote{data obtained from table 4 of \citet{Kostov14}. $m_{*,2}$ and $\delta i$ differ from \citet{Martin15}, who set $m_{*,2}=0.52 M_{\odot}$ and $\delta i = 4.02^\circ$.} 
& 0.82 & 0.54 & 0.78 & 0.48 & 0.10 & 0.037 & 0.21 & 0.40 & 0.36 & 4.07& 0.12 & 87.59 \\ \hline
Kepler 453b\footnote{data obtained from table 3 of \citet{Welsh15}. Note that $m_p$ is highly uncertain as $m_p = 0.00031\pm 0.050 M_{\rm J}$. $a_p$ differs from \citet{Martin15}, who set it to be 0.93 AU.} 
& 0.93 & 0.19 & 0.83 & 0.21 & 0.18 & 0.051 & 0.00031 & 0.56 & 0.79 & 2.30 & 0.038 & 90.28 \\ \hline
Kepler-1647b\footnote{data obtained from table 4 of \citet{Kostov15}.}
& 1.22 & 0.97 & 1.79 & 0.97 & 0.13 & 0.16 & 1.52 & 1.08 & 2.72 & 2.99 & 0.058 & 87.92 \\ \hline 
    \hline
\end{tabular}
\end{table*}
 
Note that \citet{Hinse15} has studied the possibility of a third circumbinary planet in Kepler-47 based on a single transiting event, and put an upper limit in the semi-major axis of the third planet by analyzing the transit duration. We exclude the third planet in the calculations, since the orbital parameters of this object is still largely uncertain.  

\begin{figure}[h]
\begin{center}
\includegraphics[width=3.6in]{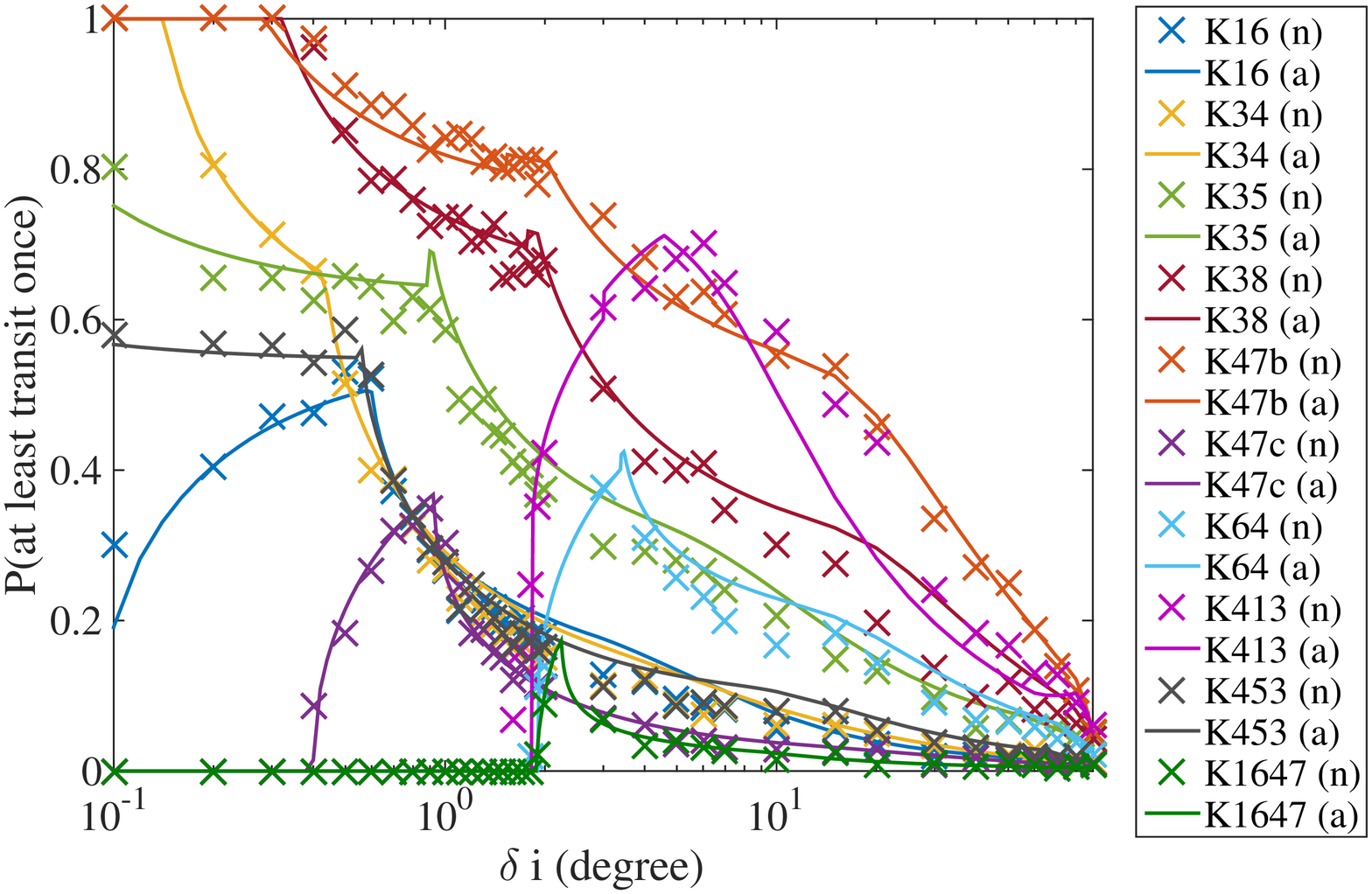} \\
\includegraphics[width=3.6in]{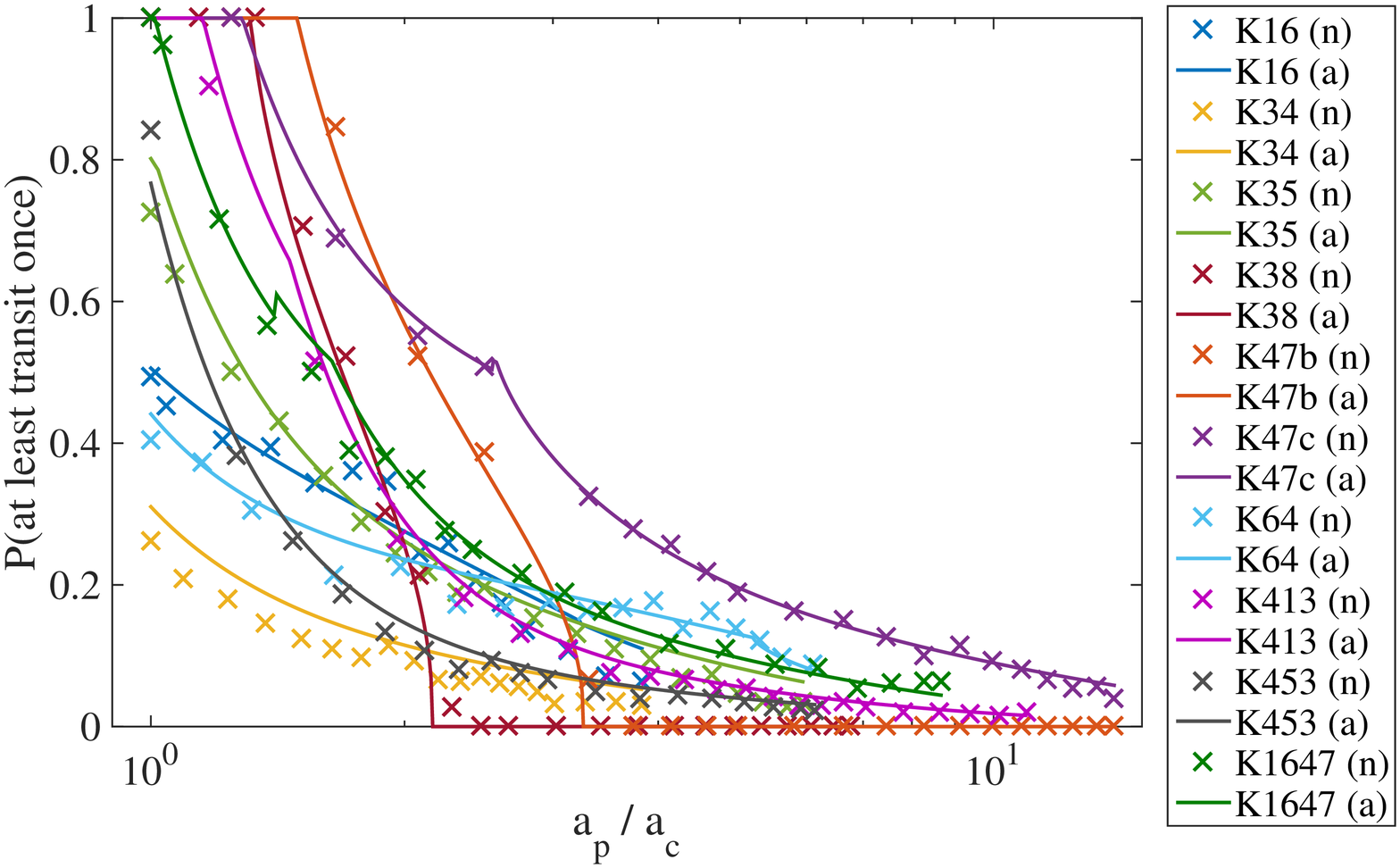} 
\caption{Probability that a planet transits at least once during the observation interval as a function of $\delta i$ (upper panel) and $a_p/a_c$ (lower panel) for the observed transiting systems. The crosses represent the numerical results and the solid lines represent the analytical results. The analytical results agree quite well with the numerical results. \label{fig:KNprob_di}}
\end{center}
\end{figure}

For each planetary systems, we use the observed properties listed in Table \ref{t:cir}, and we vary the mutual inclination between the planetary and stellar orbits or the planetary semi-major axis to obtain the transit probability as a function of the mutual inclination or the planetary semi-major axis. Then we compare the analytical results and the numerical results for different mutual inclinations and different planetary orbital semi-major axes. 

The detection periods for the different systems vary. According to the discovery papers \citep{Doyle11, Welsh12, Orosz12b, Orosz12a, Schwamb13, Kostov13, Kostov14, Welsh15, Kostov15}, the circumbinary systems are detected using different number of Kepler observation quarters, and thus, the transits occur in different total time intervals. Therefore, we set the integration time to be 600 days, 671 days, 671 days, 967 days, 1050.51 days, 967 days, 1340 days, 1470 days and 1470 days separately for Kepler-16, Kepler-34, Kepler-35, Kepler-38, Kepler-47, Kepler-64, Kepler-413, Kepler-453 and Kepler-1647 in the numerical simulations, as shown in Table \ref{t:time}. In addition, the secondary star is very faint in Kepler-38, Kepler-47, Kepler-64, Kepler-413 and Kepler-453, where only the transit of the primary star is detectable. Thus, in the numerical simulations, we only take into account the transits of the primary stars for these systems. 

\begin{table}[h]
\caption{Integration time (days) of the observed transiting cimcumbinary planets. \label{t:time} }
\centering
\begin{tabular}{ |l|l|l|l|l| }
    \hline
       Kepler 16 & Kepler-34  & Kepler-35 & Kepler-38 & Kepler-47 \\ \hline
		600 & 671 & 671 & 967 & 1050.51 \\ \hline
	   Kepler-64 & Kepler-413 & Kepler-453 & Kepler-1647 & \\ \hline
	   967 & 1340 & 1470 & 1470  & \\ \hline
    \hline
\end{tabular}
\end{table}

The upper panel of figure \ref{fig:KNprob_di} shows the probability to transit both of the stars or the primary star during the different observation interval as summarized in the paragraph above. We vary the mutual inclination ranges from $0^\circ$ to $180^\circ$. The crosses represent the results from numerical simulations, and the solid lines represent the analytical results. It is shown that the analytical results are consistent with the numerical results for both the case considering the transit of both stars and the transit of the primary stars. The planet still has a high probability to transit when the mutual inclination reaches $\sim 5^\circ$. Note that at high inclinations, the ascending node librates, and this may introduce the discrepancy between the numerical results and the analytical results, as discussed by \citet{Martin15}. In addition, planets with higher mutual inclination ($\sim1-3^\circ$) may be more likely to transit, for systems with large $|\Delta i_b|$, as pointed out by \citet{Martin15}.

The lower panel of figure \ref{fig:KNprob_di} shows the probability to transit at least once in the observational interval as a function of the planetary semi-major axis to the stability ratio. We set the minimum planetary semi-major axis to be the critical semi-major axis beyond which the planet is stable (from equation \ref{eqn:sta}) and we set the maximum semi-major axis to be that corresponding to a four year orbit. Overall, the analytical results (solid lines) are also consistent with the numerical results (crosses). The numerical probability can be used to derive many properties of the architecture of the circumbinaries in the next section (\textsection \ref{s:arc}).

\section{Circumbinary Planetary Architecture}
\label{s:arc}
The architectures of the circumbinary planetary systems provide important clues on the formation of planetary systems. In this section, we focus on the observed transiting circumbinary systems and study their orbital properties. To accurately determine the role of the selection bias, we use the transit probability from numerical integrations directly. Note that in addition to transits, another indicator of circumbinary planets is a variation in the eclipse timings (ETVs). This is noticeable in roughly half of the Kepler sample, and may introduce a detection bias which we do not consider. This approach differs from \citet{Armstrong14} and \citet{Martin14}, who studied the abundance of circumbinary planetary systems using population synthesis.

\subsection{Distribution of $a_p$}
\label{s:ap}
It has been found that most of the innermost transiting circumbinary planets reside near the stability limit close to the stellar binary \citep{Armstrong14}. This may indicate the dominance of migration during planet formation. However, this may be also due to selection effects, because close-in planets admit larger orbital parameter spaces which allow transits, and thus are more likely to be detected. Using population synthesis, \citet{Martin14} found that selection biases alone cannot account for the pile up near the stability limit. Recently, Kepler-1647 was discovered to orbit far from the stability limit \citep{Kostov15}. In this section, we consider multiple semi-major axis distributions and include the newly discovered Kepler-1647 to study the pile up of planets near the stability limit, using a Bayesian approach to take into account the selection bias. 

\begin{figure}[h]
\begin{center}
\includegraphics[width=3.3in]{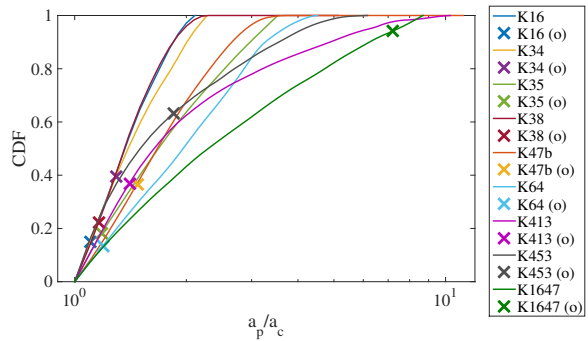} 
\caption{Cumulative distribution function of the scaled semi-major axis ($\tilde{a}_p = a_p/a_c$) given that the planet transit at least twice. The crosses represent the observed value of the innermost planets for each system. Except Kepler-1647b and Kepler 453b, the probability that the scaled semi-major is smaller than the observed value is $\lesssim 50\%$. \label{fig:KNprob_5}}
\end{center}
\end{figure}

To take into account selection effects, we require planets to transit at least twice in order to be detected. Then, we study the significance of the pile-up using a hypothesis test. Specifically, our null hypothesis is that the distribution of the detected planetary semi-major axis follows the conditional probability distribution of the semi-major axis given that the planets transit at least twice ($P(\tilde{a}_p | tt_2)$), where $tt_2$ stands for the event that a planet transits at least twice. Then, we calculate the probability that the planetary semi-major axis is smaller than the observed value. If this probability is very small, it rejects the null hypothesis and indicates that the reason the planet locates near the stability limit is not only due to selection effects. $P(\tilde{a}_p | tt_2)$ ($\tilde{a}_p = a_p/a_c$) using the Bayesian approach can be expressed as the following:
\begin{align}
P(\tilde{a}_p | tt_2) = \frac{P(tt_2 | \tilde{a}_p) P(\tilde{a}_p)}{\int_{(\tilde{a}_p)_{min}}^{(\tilde{a}_p)_{max}} P(tt_2 | a') P(a') ~da'} ,
\label{eqn:probap}
\end{align}
where $\tilde{a}_{p, min} = 1$ for stability purposes, and $\tilde{a}_{p, max}$ corresponds to orbital period equals to the total time of detection obtained from the discovery papers, as summarized in the beginning of this section \textsection \ref{s:arc}. 

$P(tt_2 | \tilde{a}_p)$ stands for the probability to transit at least twice at different planetary semi-major axis $\tilde{a}_p$. This probability can be obtained using the analytical expression or using numerical simulations as described in the previous section. We use the numerical values directly for the following analysis. $P(\tilde{a}_p)$ is the prior of $\tilde{a}_p$, and we assume a uniform distribution for the following reasons. First, the signal to noise level is important in the detection of the circumbinary planets. In particular, the signal to noise level ($s/n$) of the transit depends on the distance between the planet and the stellar binary: 
\begin{equation}
s/n \propto \sqrt{n_{tr} t_{dur}} ,
\end{equation}
where $n_{tr}$ stands for the number of transits, $t_{dur}$ is the transit duration time. The transits of the same circumbinary system can be very different depending on the relative velocity between the star and the planet during the transits, and thus each transit needs to be resolved separately. Therefore, $s/n \propto \sqrt{t_{dur}}$. The explicit expression for the transit has been derived by \citet{Kostov14}, where the dependence on $a_p$ is weak, and the duration time increases when the planet-star distance increases assuming the impact parameter is independent of $a_p$, since the dependence of the impact parameter on $a_p$ is not trivial especially when $\Delta_{ib} \ne 0$. Thus, it is easier to detect the planet when the planet is farther away in terms of this signal to noise level. To obtain the lower limit constraint (maximum value of $P(\tilde{a}_p <  \tilde{a}_{p, obs}| tt_2)$) on the pile up near the stability limit, we use a uniform prior where $P(\tilde{a}_p) = 1/((\tilde{a}_p)_{max}-(\tilde{a}_p)_{min})$. Secondly, note that \citet{Armstrong14} simulated the recovery rate of transit detection for circumbinary systems and showed that the recovery rate decreases mildly with orbital period, based on simulations with $P_p = 10.2P_b$ and $P_p = 300$ days. However, a detailed scaling was not included. The decrease of the detection probability as a function of planet distance may be inherited in the detection algorithm, where a larger number of the transits makes it less likely to miss the transits. Since the recovery rate only decreases mildly and no detailed scaling as a function of $a_p$ is available yet, we do not take this into account here. 

The cumulative distribution of $\tilde{a}_p$ given the planet transits twice for the observed innermost planet is shown in figure \ref{fig:KNprob_5}. The crosses represent the observed results. Except the newly discovered Kepler-453b and Kepler-1647b, which have large probability that $\tilde{a}_p$ is smaller than the observed value, most of the planets are moderately close to the stability limit, with probability $\lesssim 40\%$. However, these probabilities are not small enough to reject the null hypothesis.

\begin{figure}[h]
\begin{center}
\includegraphics[width=3.3in]{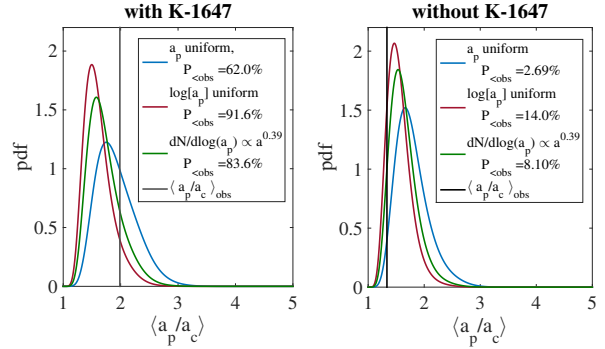} 
\caption{Probability distribution function of the mean of $a_p/a_c$ of the innermost planets in the observed transiting systems. The left panel represents the case with Kepler-1647b, and the right panel represent the case without Kepler-1647b. The solid black lines indicate the observed values. The different colored lines represent the case with different prior distribution. Excluding Kepler-1647b, the probability that $a_p/a_c$ is smaller than the observed value is very small, indicating that the pile-up near the stability limit is not due to selection effect if the prior is uniform in $a_p/a_c$. \label{fig:convolve}}
\end{center}
\end{figure}

Since most of the innermost planets (except Kepler-1647b) have $P(\tilde{a}_p < \tilde{a}_{p, obs}| tt_2) < 50\%$, the collective feature of these systems may suggest that there exists a pile-up near the stability limit. To investigate the collective behavior, we designed a numerical hypothesis test. Specifically, we take all the observed transiting circumbinary systems together, and use the averaged $\tilde{a}_p$ ($\langle \tilde{a}_p \rangle$) as a statistic to test the null hypothesis that $\tilde{a}_p$ follows $P(\tilde{a}_p | tt_2)$ according to equation (\ref{eqn:probap}) for each system. Specifically, if under the null hypothesis, the observed $\langle \tilde{a}_p \rangle$ or the values smaller than that has a very small probability ($<5\%$), we reject the null hypothesis, and we claim that there is likely a pile-up of planets near the stability after taking into account selection effects. 

We numerically convolve the distribution of $\tilde{a}_p$ for all the systems in order to obtain the distribution of $\langle \tilde{a}_p \rangle$, and the result is represented by the blue lines in figure \ref{fig:convolve}. Excluding Kepler-1647, the probability that the averaged $\tilde{a}_p$ is smaller than the observed value is very small ($2.69\%$), suggesting that there is likely a pile-up after considering the selection effects. This is consistent with the population synthesis study by \citet{Martin14}. However, including Kepler-1647, the probability is much larger (reaching $\sim 62\%$) indicating that the null hypothesis cannot be rejected. There are two possibilities: if Kepler-1647 shares the same distribution as the other nine systems, there is likely no pile-up of planets near the stability limit; if Kepler-1647 is an outlier of this sample, which follows a different semi-major axis distribution, then there is likely a pile-up for some population of the circumbinary planetary systems. More observations of the transiting circumbinary systems can help distinguish this. Note that the probability only differs within a factor of two if we take transits of both stars into account and integrate over four years for all the observed systems.

The distribution of planetary periods for single star systems has been studied in the literature \citep{Winn15}. For instance, for small size planets ($1-4 {\rm R}_{\oplus}$) with period range of $20-200$ days, \citet{Silburt14} have found that the planetary period follows a log-uniform distribution, where $dN/da_p \sim \propto a_p^{-1}$, consistent with \citep{Youdin11, Howard12, Petigura13, Fressin13}. For larger size planets ($4-8 {\rm R}_{\oplus}$), the probability density can be expressed as $dN/dlogP_p \propto P_p^{0.7}$ \citep{Dong13}, where the semi-major axis distribution is nearly uniform $dN/da_p \sim \propto a_p^{0}$. From radio velocity studies, \citet{Cumming08} obtained that $dN/dlogP_p \propto P_p^{0.26}$, where $dN/da_p \sim \propto a_p^{-0.61}$ for planet mass $> 0.4 M_{J}$, and orbital period $<2000$ days. Next, we check that whether the circumbinary planetary systems may follow similar distributions as the planets around single stars, and whether this in addition to the selection effect can explain the observed pile-up.

Using a log-uniform distribution as a prior, the results on the probability density function of $\langle \tilde{a}_p \rangle$ are shown by the red lines in figure \ref{fig:convolve}. Excluding Kepler-1647, the probability that $\langle \tilde{a}_p \rangle$ is smaller than the observed value is $14\%$, and including Kepler-1647, the probability that $\langle \tilde{a}_p \rangle$ is smaller than the observed value $91.6\%$. Both cases cannot rule out the hypothesis that the planetary period follows a log-uniform distribution, suggesting that there is no additional pile-up if the circumbinary planets share the log-uniform period distribution as the small size planets around single stars. The green lines in figure \ref{fig:convolve} shows the case when the prior follows $dN/d\tilde{a}_p \sim \propto \tilde{a}_p^{-0.61}$, the probability that $\langle \tilde{a}_p \rangle$ is smaller than the observed value is 8.1\% excluding Kepler-1647, and is 83.6\% including Kepler-1647. Neither of the cases rule out the hypothesis that the circumbinary planetary system follows the similar distribution ($dN/dlogP_p \propto P_p^{0.26}$) as the planets around single stars obtained from the RV measurements by \citet{Cumming08}. This also suggests that the pile-up is consistent with this period distribution and selection effects. On the other hand, the circumbinary planets do not favor the period distribution of large planets around the single stars ($dN/dlogP_p \propto P_p^{0.7}$) obtained by \citet{Dong13}, where selection effects alone cannot explain the pile-up near the stability limit.

\subsection{Coplanarity}
\label{s:coplanarity}
The observed transiting circumbinary planets all have small mutual inclinations between their planetary orbits and the stellar binary (as shown in Table \ref{t:cir}). This may be primordial since the observed circumbinary protoplanetary disks are also aligned with the stellar orbit within $\sim 3^\circ$ \citep[e.g.,][]{Czekala16}. However, this may also be due to selection effects, because systems with near coplanar configurations are more likely to be observed via the transit method. To test whether the coplanarity is only a selection effect, and to put a constraint on the mutual inclination distribution, we identify the probability distribution of the mutual inclination that is consistent with the observations while taking into account the selection bias. 

Similar to our study on the distribution of planetary semi-major axis in the previous section, we require the planet to transit at least twice for a robust detection, and our null hypothesis is that the distribution of the observed mutual inclination follows the conditional probability distribution given that the planet transits at least twice ($P(\delta i|tt_2)$), where $\delta i$ is the mutual inclination, $tt_2$ represent the event that a planet transits at least twice. If the probability that the mutual inclination is smaller than the observed value is very small ($<5\%$), it rejects the null hypothesis and it indicates that the mutual inclination follows a distribution with a smaller spread than the prior. Specifically, 
\begin{align}
P(\delta i|tt_2) = \frac{P(tt_2|\delta i)P(\delta i)}{\int_{i'=0}^{i'=180} P(tt_2|\delta i = i')P(\delta i = i') di'} ,
\label{eqn:iprob}
\end{align}
where $P(tt_2|\delta i)$ can be obtained from the analytical approach. In the following analysis, we directly use results from the numerical simulations as described in section \textsection \ref{s:usb}.

Assuming an isotropic distribution as the prior ($P(\delta i)= \sin{i}/2$), the cumulative distribution of $P(\delta i|tt_2)$ is shown in figure \ref{fig:KNprob_di}, where the crosses represent the observed mutual inclination. It shows that the probability that the mutual inclination is smaller than the observed value is very small ($\lesssim 1\%$). Thus, it is highly unlikely that the observed coplanarity of the systems is only due to selection effects. This suggests that the observed circumbinary planets are likely formed near the orbital plane of the stellar binary.

\begin{figure}
\begin{center}
\includegraphics[width=3.3in]{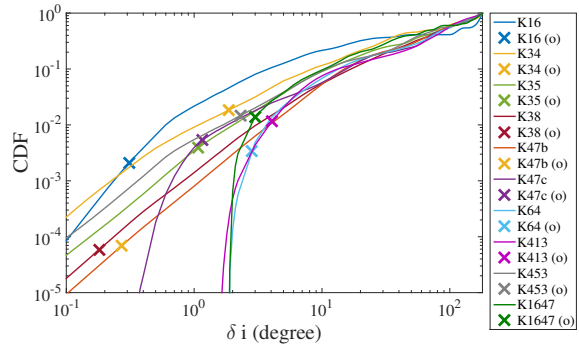} 
\caption{Probability of the mutual inclination given that the planet transits at least twice, assuming the prior distribution of $\delta i$ is isotropic. The crosses represent the observed value. The probability that the mutual inclination is smaller than the observed value is very small, indicating that the mutual inclination is likely small. \label{fig:KNprob_di}}
\end{center}
\end{figure}

We next use different prior distributions to further investigate the distribution of the mutual inclination. We assume that the prior of the mutual inclination follows a Fisher distribution ($f(\delta i | \kappa)$), also known as a $p=3$ von Mises-Fisher distribution, which is a probability distribution on the 2-dimensional sphere in the 3-dimensional space. This is similar to the model of the spin-orbit misalignment distribution discussed in the literature \citep[e.g.,][]{FabryckyWinn09, Li15}. Specifically,
\begin{equation}
 f_{\kappa}(\delta i) = \frac{\kappa}{2\sinh{\kappa}} e^{\kappa \cos{\delta i}} \sin{\delta i}, \label{eqn:fish}
\end{equation}
where the concentration parameter $\kappa$ controls the spread in mutual inclination. For large $\kappa$, $f_{\kappa}(\delta i)$ approaches Rayleigh distribution with width $\sigma \to \kappa^{-1/2}$, and when $\kappa \to 0$, the distribution is isotropic.  

\begin{figure}[h]
\begin{center}
\includegraphics[width=3.3in]{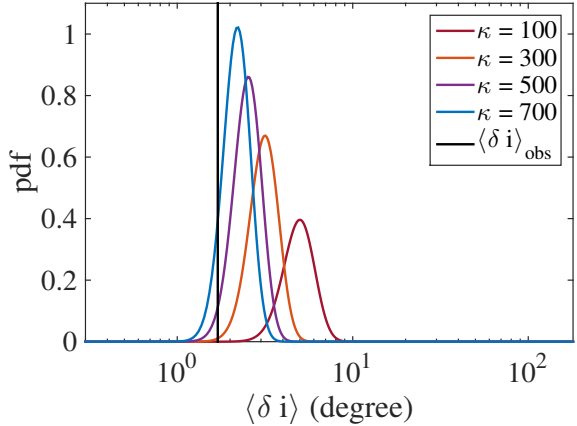} 
\caption{Probability distribution of the averaged mutual inclination with different prior distribution of $\delta i$. The solid black line indicates the observed value. \label{fig:convolve_di}}
\end{center}
\end{figure}

For large $\kappa$, the prior distribution of the inclination concentrates in the near co-planar regime, and the probability of the mutual inclination to be smaller than the observed values may be $\lesssim50\%$ for many of the observed systems. Then, the collective behavior of the observed systems may still indicate a narrower spread. Therefore, similar to our study on the semi-major axis, we design numerical hypothesis tests and use the average mutual inclination as a statistic to select the distribution that fits well with the observation. Specifically, the null hypothesis is that $\delta i$ follows the distribution of $\delta i$ according to the conditional probability in equation (\ref{eqn:iprob}) for each system, and the null hypothesis can be rejected if the observed average $\delta i$ or values smaller than that has a very small probability ($<5\%$) under the null hypothesis.

We include four prior distributions with four different $\kappa$: $\kappa = 100$, $\kappa = 300$, $\kappa = 500$ and $\kappa = 700$, and we calculated the convolved distribution of the observed systems to obtain the distribution of $\langle \delta i \rangle$. The results are shown in figure \ref{fig:convolve_di}. The solid black line indicates the observed averaged mutual inclination. For the four prior distributions, the averaged mutual inclinations are $7.2^\circ$, $4.1^\circ$, $3.2^\circ$ and $2.7^\circ$, and the standard deviations are $3.8^\circ$, $2.2^\circ$, $1.7^\circ$ and $1.4^\circ$. The probability that the average mutual inclination is smaller than the observed value is $5\times10^{-4}\%$, $0.19\%$, $2.2\%$ and $8.8\%$ for $\kappa = 100$, $\kappa = 300$, $\kappa = 500$ and $\kappa = 700$ respectively. Thus, the hypothesis can be rejected when $\kappa = 100$, $\kappa = 300$ and $\kappa = 500$. In addition, it indicates that the mutual inclination distribution is more consistent with the observation for $\kappa > 500$, corresponding to an average mutual inclination of $\lesssim 3^\circ$. 

The near co-planar ($\lesssim 3^\circ$) feature of the circumbinary planetary system is consistent with the coplanarity of the multi-transiting planetary systems with a single star (multis), where the study of transit duration ratios \citep{Fang12, Fabrycky14} and population synthesis studies \citep{Ballard16, Moriarty15} suggest that most of the multis have mutual orbital inclinations less than $\sim 3^\circ$. Moreover, the observed circumbinary protoplanetary disks are also quite aligned with the stellar orbits ($\lesssim 3^\circ$) \citep{Rosenfeld12, Czekala15, Czekala16}, and this may indicate that the coplanarity of the circumbinary planets are primordial. In addition, based on the abundance studies by \citet{Armstrong14, Martin14}, the co-planarity of the circumbinary systems may indicate that the occurrence rate of the circumbinary systems is similar to that of the single star systems.

\subsection{Multis vs. Singles}
Although nine out of the ten observed transiting circumbinary systems are single-transiting systems, it does not necessarily mean that circumbinary systems more likely contain a single planet, because farther companions are more difficult to detect via the transit method. In this section, we take into account the selection effects and investigate the multiplicity and planet-planet spacing of the planetary systems.

The transit probability of the outer companion is sensitive to its location, as the transit probability decreases with star-planet separation. On the other hand, outer companions of the circumbinary planets cannot be located very close to the inner planets, because closely separated of planets are unstable due to the planet-planet interactions. It has been found that the observed spacing of the Kepler systems is clustered around $\sim 12$ mutual Hill radii ($R_H$), and it coincides with the required spacing for stability obtained using N-body simulations \citep[e.g.,][]{Pu15}. Dynamics of multi-planet circumbinary systems has been investigated by \citet{Kratter14, Smullen16}. In particular, it was found that the intra-planet spacing is of order $5-7 R_H$ when the inner planet is close to $a_c$, and the spacing of the planet for the binary case asymptotes to the single star results when the inner planet is farther ($a_p \sim 1.5-2 a_c$). For simplicity, we mark the location of the outer companion at $12 R_H$ for illustration. The mutual Hill radius of the single stellar system is expressed as $R_{H, single} = (a_1+a_2)/2 \times ((m_{p,1}+m_{p,2})/(3M_*))^{1/3}$, where $a_1$ \& $a_2$ and $m_{p,1}$ \& $m_{p,2}$ are the semi-major axes and the masses of the planets and $M_*$ is the mass of the host star. For the circumstellar system, we set the mutual Hill radius to be:
\begin{equation}
R_H = \frac{a_1+a_2}{2} \Big[\frac{m_{p,1}+m_{p,2}}{3(m_1+m_2)}\Big]^{1/3} .
\end{equation}
Since the mass of the companion planet is not known, we set the companion planet to be a test particle with mass zero to obtain the maximum transit probability when the planet separation is the smallest. We set the companion planet mass to be two Jupiter mass to probe the minimum probability when the planet separation is larger.

\begin{figure*}[h]
\begin{center}
\includegraphics[width=6.6in]{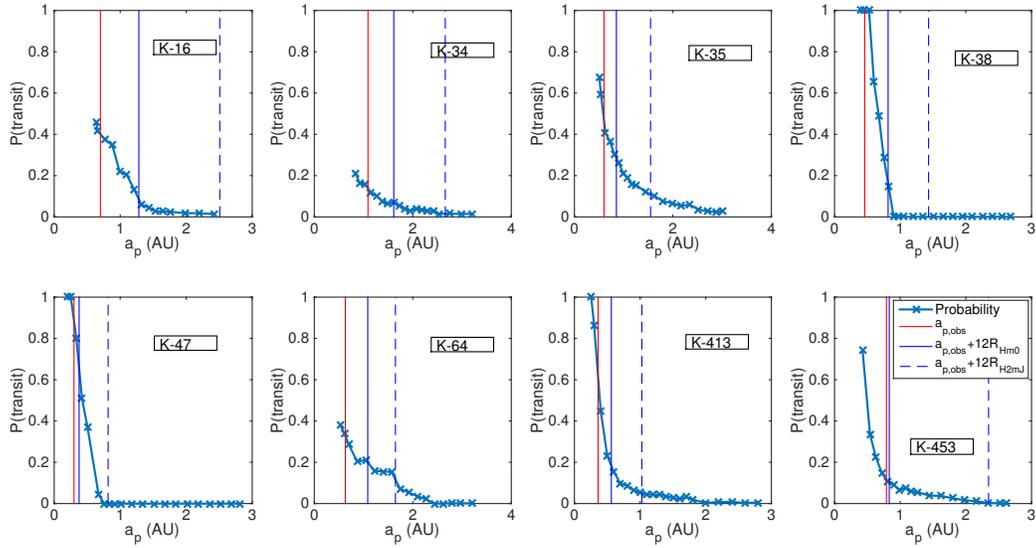} 
\caption{Probability that a planet transits at least twice as a function of $a_p$ for the observed systems. The red line indicates the location of the observed innermost planet. The solid blue line indicates the location of a companion test particle at 12 mutual Hill radii away from the innermost planet, and the dashed blue line indicates the location when the mass of the companion is two Jupiter masses. \label{fig:multi}}
\end{center}
\end{figure*}

Similar to the previous sections, we require the planet to transit at least twice for a robust detection criterion. Then, we calculate numerically the probability to transit at least twice at different semi-major axes for the observed systems, assuming the the companion planet share the same mutual inclination with the innermost planet. A small probability at $a_p \gtrsim 12R_H$ implies that it is unlikely to detect the companion, and thus it is possible to have farther undetected companions in the system. 

The results are shown in figure \ref{fig:multi}. The solid red line indicates the semi-major axis of the observed planet, and the solid (dashed) blue line represents the semi-major axis at 12 $R_H$ away from the detected planet, assuming the companion planet has mass zero (2 Jupiter masses). Note that for Kepler-1647, the planet orbital period is longer than 1470 days ($\sim$ four years) at twelve mutual Hill radii away even when the companion is a test particle, so the probability to transit at least twice is zero. Thus, we exclude Kepler-1647 in the figure. Figure \ref{fig:multi} shows that the probability to detect the outer companion is quite low, except for Kepler-47 if the planet mass is low. It is consistent with the observation where Kepler-47 indeed has multiple planets. Thus, we find no strong evidence that the circumbinary systems more likely contain a single planet.

\subsection{Stellar Binary Period}

It has been shown that the observed transiting circumbinary planets orbit around stellar binaries with long orbital periods ($\gtrsim 7$ days) \citep[e.g.,][]{Armstrong14, Martin14}. However, a large number of eclipsing binaries have short orbital periods ($\lesssim 3$ day) \citep[e.g.,][]{Slawson11}. The absence of the circumbinary systems may indicate that it is difficult to form planets around short period binaries. In addition, it can also be caused by Lidov-Kozai mechanism, which contributes to the formation of short period binaries. Specifically, the short period stellar binaries are formed through the Lidov-Kozai mechanism, and their inclination and eccentricity oscillate due to the perturbation of a third companion \citep{Mazeh79, Fabrycky07}. Note that the planet does not cause Lidov-Kozai oscillations in the stellar binary since it is not massive enough, as studied by \citet{Migaszewski11, Martin16}, where the third companion which produces the short period binaries needs to be massive. During this process, planets can be ejected or collide with the star, and the survived planets end up with inclined orbits with respect to the stellar orbit to avoid transits \citep{Munoz15, Martin15b, Hamers16}. However, with precession, the transit probability at high mutual inclination can still be large. In this section, we study the probability distribution of the stellar binary orbital period including the misaligned cases. 

First, we use the analytical result from \textsection \ref{s:tprob} to obtain the transit probability, where for simplicity we set the stellar properties to those of the Sun, and we set the stellar binary to be aligned with the line of sight. Next, we set the prior period distribution of the eclipsing binary to be that of the {\it Kepler} sample. We integrate the probability over the mutual inclination and obtain the probability distribution of $P_b$ joint with the event that the planet transits at least once. Specifically,
\begin{align}
P(P_{b} \cap tt_1) &= P(P_{b}) \Big(\int P(tt_1|\delta i', P_{b}) P(\delta i')~d\delta i' \Big) ,
\end{align}
where $tt_1$ represent the event that the planet transits at least once in four years. 

\begin{figure}[h]
\begin{center}
\includegraphics[width=3.3in]{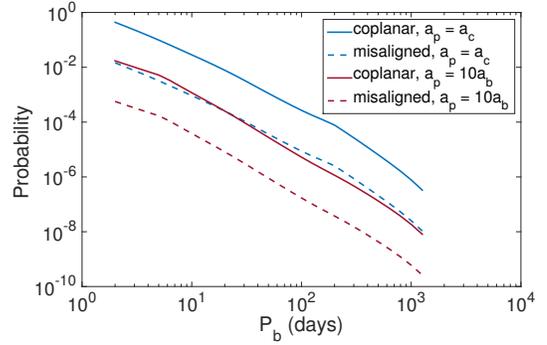} 
\caption{Probability to transit at least once for different stellar binary period ($P_b$). The solid lines represent the near coplanar case and the dashed lines represent the highly misaligned case. The blue color represent the case when the planet is at the stability limit, and the red color represent the case when $a_p = 10 a_b$, motivated by the Lidov-Kozai mechanism. The misaligned short period stellar binary transit probability is similar to that of the aligned aligned case when the planet is close to the star, but the probability decreases when the planet is farther. \label{fig:Pl5}}
\end{center}
\end{figure}

To compare the coplanar and the misaligned cases, we include a near coplanar distribution $\delta i \lesssim 3^\circ$, as discussed in the previous section \textsection \ref{s:coplanarity}, and a highly misaligned distribution $\delta i \in [40^\circ, 140^\circ]$, motivated by the Lidov-Kozai formation mechanism. For the case when the mutual inclination is less than $3^\circ$, we use the Fisher distribution with $\kappa = 500$, and for the case when the mutual inclination is high, we set the distribution to be $P(\delta i) \propto \sin(\delta i)$, with lower and upper bound to be $40^\circ$ and $140^\circ$ separately. 

Figure \ref{fig:Pl5} shows the result. The solid lines represent the coplanar case, and the dashed lines represent the misaligned case. In addition, the blue lines indicate the case that $a_p/a_c = 1$, and the purple lines indicates the case when $a_p/a_b = 10$ motivated by the Lidov-Kozai mechanism, where the inner binaries shrink during the formation of the short period systems and thus the semi-major axis ratio of the planet to the stellar binary increases. We set the minimum $P_b$ to be 2 days because the signal to noise level is lower when the stellar binary orbital period is shorter, and we set the maximum $P_b$ to be four years. A detailed study on the short period limit due to the signal to noise level is important, but is beyond the scope of this article. Note that since we set the stellar binary to be aligned with the line of sight, the actual transit probability for the coplanar case should be moderately lower than the results shown in figure \ref{fig:Pl5} when $\Delta_{ib}\ne0$.

Taking into account the abundance of the short period binaries, the transit probability for the high mutual inclination short period stellar binary is similar to that of the aligned long stellar period case, when the planets locate near the stability limit. However, the planetary to stellar semi-major axis ratio increases during the formation mechanism through Lidov-Kozai oscillations, as the stellar binary orbit shrinks. The increase of $a_p/a_b$ further reduces the transit probability. Therefore, the circumbinary planets around short period stellar binaries are still unlikely to be detected through the transit method. In other words, the formation mechanisms involving the Lidov-Kozai mechanism are consistent with the observations. This also implies that the planets likely do not move closer to the stellar binaries after the formation of the short period stellar binaries. 

Although using transit methods it is unlikely to detect the misaligned circumbinary planets at far distances from the short period stellar binaries, these planets can be detected through the eclipsing timing variation method. As the center mass of the stellar binary moves around the barycenter of the system, it causes variations in the light travel time from the stellar binary to the observer \citep[e.g.,][]{Schneider95, Schwarz11}. For hot Jupiters orbiting around solar type stellar binaries at 1 AU, this effect causes a time variation of the eclipses at the scale of 1 second, and this is detectable using {\it Kepler} for a 9 magnitude target \citep{Sybilsky10}. This effect is stronger when the stellar mass is lower, and when the planets are farther away yet with periods shorter than the observation time. 

\section{Conclusions}
In this paper, we investigate the architectural properties of the planetary systems corrected by selection effects. First, we revisit the planetary stability limit when the planetary orbit is misaligned with the stellar binary. We find that the system is more stable when the mutual inclination is higher, which is consistent with \citet{Doolin11}, and we find that the variations in the semi-major axes of the planets show interesting patterns. Next, we derive the analytical expression for the transit probability in a realistic setting, where a finite observation period and planetary orbital precession are both included. The analytical results agree well with the numerical simulations. In particular, the probability to transit one of the binary stars is shown in equation (\ref{eqn:tprob1}), and the probability to transit both stars is shown in equation (\ref{eqn:tprob}). Different from the case with infinite observation time period \citep{Martin15}, the transit probability does not always increase as a function of mutual inclination (as shown in figure 7). In addition, comparing the transit probabilities of the circumbinary systems and systems with a single star, the transit probability for circumbinary systems can be lower if the observation period is very short (e.g., $\sim 30$ days, when $P_b = 5$ days, $a_p/a_b =5$, and $\delta i \gtrsim 50^\circ$). Thus, the transit probability of some circumbinary planets may be lower than their single star counterparts for the {\it TESS} mission, especially when the mutual inclination is high. On the other hand, the transit probability for the circumbinary planets is likely higher for the {\it K2, PLATO, Kepler} missions.

Using the transit probability, we obtain architectural properties of the circumbinary systems. First, we study the distribution of planetary semi-major axis. Nine out of the ten observed circumbinary systems host innermost planets moderately close to the stability limit. However, the ninth system (Kepler-1647) hosts a planet that is much farther from the stability limit. Assuming that the tenth system is from a different distribution, there is only a small probability that the pile up of planets near the stability limit is due to selection bias for the nine systems . This implies the dominance of migration during planet formation for a population of the circumbinary planetary systems. On the other hand, assuming that Kepler-1647 is in the same distribution, then, there is no strong evidence for a pile up of planets near the stability limit. Observations of more circumbinary planetary systems can help distinguish these two scenarios. Moreover, we find that the pile-up is consistent with a log-uniform distribution of the planetary semi-major axis.

We next study the distribution of the mutual inclination between the planetary orbits and the orbits of the stellar binaries. All of the observed circumbinary planets to date are near coplanar with the stellar orbits. The mutual inclination between the planet orbit and the stellar binary is much smaller than the result of an isotropic distribution after taking into account the selection effects. We find that the mutual inclination can be fit well with a Fisher distribution of $\kappa \gtrsim 500$, corresponding to a average mutual inclination of $\lesssim 3^\circ$. This is similar to the mutual inclination for the multi-transiting systems around single stars \citep{Fang12, Fabrycky14, Ballard16, Moriarty15}. Since the circumbinary protoplanetary disk also align with the stellar orbits within $\sim 3^\circ$ \citep{Rosenfeld12, Czekala15, Czekala16}, this may indicate a primordial alignment of the circumbinary planetary orbits.

Current observation seems to suggest that only one out of the ten observed circumbinary planetary systems hosts multiple planets. This can either be a result of the selection effects or imply that circumbinary planetary systems tend to host a single planet. To investigate this, we find that the probability to detect outer companion is very small for most of the systems, assuming a separation of $\sim 12 R_H$ for stability purposes. Thus, we do not find strong evidence that the circumbinary planetary systems preferentially host a single planet. This indicates that the observed systems may have outer companions, but it is difficult to detect them. 

Finally, we investigate the transit probability of systems with short period stellar binaries and with inclined planetary orbits, motivated by the lack of observed circumbinary planets around short period stellar binaries. We find that considering the period distribution of eclipsing binaries, the transit probability of the misaligned system is similar to that of the aligned long stellar period systems if the planet is located near the stability limit. However, the transit probability decreases as the planetary to stellar semi-major axis ratio decreases. This shows that the observation is consistent with the formation mechanism involving Lidov-Kozai oscillation, where the mutual inclination is excited and the semi-major axis ratio is reduced if the planets survives during the formation process \citep{Munoz15, Martin15b, Hamers16}. It also implies that the planets do not move closer to the stellar binary after the misalignment. Instead of transit methods, eclipsing time variation may provide a way to detect such misaligned large semi-major axis ratio circumbinary planets.

\section*{Acknowledgments}
The authors would like to thank the anonymous referee for giving constructive comments, which substantially improved the quality of this paper. In addition, the authors would like to thank Yanqin Wu and Josh Winn for helpful discussions. This work and MT were supported in part by NSF DMS-1521667.  GL was supported in part by Harvard William F. Milton Award.




\bibliographystyle{hapj}
\bibliography{msref}

\end{document}